\documentstyle[epsfig]{mn}

\newif\ifAMStwofonts
                                                                                                                                                     
\ifoldfss
  \ifCUPmtlplainloaded \else
    \NewTextAlphabet{textbfit} {cmbxti10} {}
    \NewTextAlphabet{textbfss} {cmssbx10} {}
    \NewMathAlphabet{mathbfit} {cmbxti10} {} 
    \NewMathAlphabet{mathbfss} {cmssbx10} {} 
  \fi
  \ifAMStwofonts
    \ifCUPmtlplainloaded \else
      \NewSymbolFont{upmath} {eurm10}
      \NewSymbolFont{AMSa} {msam10}
      \NewMathSymbol{\upi}     {0}{upmath}{19}
      \NewMathSymbol{\umu}     {0}{upmath}{16}
      \NewMathSymbol{\upartial}{0}{upmath}{40}
      \NewMathSymbol{\leqslant}{3}{AMSa}{36}
      \NewMathSymbol{\geqslant}{3}{AMSa}{3E}

    \fi
  \fi
\fi 
\ifnfssone
  \newmathalphabet{\mathit}
  \addtoversion{normal}{\mathit}{cmr}{m}{it}
  \addtoversion{bold}{\mathit}{cmr}{bx}{it}
  \newmathalphabet{\mathbfit} 
  \addtoversion{normal}{\mathbfit}{cmr}{bx}{it}
  \addtoversion{bold}{\mathbfit}{cmr}{bx}{it}
  \newmathalphabet{\mathbfss} 
  \addtoversion{normal}{\mathbfss}{cmss}{bx}{n}
  \addtoversion{bold}{\mathbfss}{cmss}{bx}{n}
  \ifAMStwofonts
    \ifCUPmtlplainloaded \else
      \UseAMStwoboldmath
      \makeatletter
      \new@mathgroup\upmath@group
      \define@mathgroup\mv@normal\upmath@group{eur}{m}{n}
      \define@mathgroup\mv@bold\upmath@group{eur}{b}{n}
      \edef\UPM{\hexnumber\upmath@group}
      \new@mathgroup\amsa@group
      \define@mathgroup\mv@normal\amsa@group{msa}{m}{n}
      \define@mathgroup\mv@bold\amsa@group{msa}{m}{n}
      \edef\AMSa{\hexnumber\amsa@group}
      \makeatother
      \mathchardef\upi=''0\UPM19
      \mathchardef\umu=''0\UPM16
      \mathchardef\upartial=''0\UPM40
      \mathchardef\leqslant=''3\AMSa36
      \mathchardef\geqslant=''3\AMSa3E
    \fi
  \fi
\fi 
                                                                                                                                                     
\ifnfsstwo
  \DeclareMathAlphabet{\mathbfit}{OT1}{cmr}{bx}{it}
  \SetMathAlphabet\mathbfit{bold}{OT1}{cmr}{bx}{it}
  \DeclareMathAlphabet{\mathbfss}{OT1}{cmss}{bx}{n}
  \SetMathAlphabet\mathbfss{bold}{OT1}{cmss}{bx}{n}
  \ifAMStwofonts
    \ifCUPmtlplainloaded \else
      \DeclareSymbolFont{UPM}{U}{eur}{m}{n}
      \SetSymbolFont{UPM}{bold}{U}{eur}{b}{n}
      \DeclareSymbolFont{AMSa}{U}{msa}{m}{n}
      \DeclareMathSymbol{\upi}{0}{UPM}{``19}
      \DeclareMathSymbol{\umu}{0}{UPM}{``16}
      \DeclareMathSymbol{\upartial}{0}{UPM}{``40}
      \DeclareMathSymbol{\leqslant}{3}{AMSa}{``36}
      \DeclareMathSymbol{\geqslant}{3}{AMSa}{``3E}
    \fi
  \fi
\fi 
                                                                                                                                                     
\ifCUPmtlplainloaded \else
  \ifAMStwofonts \else 
    \def\upi{\pi}
    \def\umu{\mu}
    \def\upartial{\partial}
  \fi
\fi

\title[A non-Gaussian Spot in WMAP]{Detection of a non-Gaussian Spot in WMAP}

\author[M. Cruz et al.]{M. Cruz,$^{1,2}$\thanks{e-mail:
cruz@ifca.unican.es}
E. Mart\'{\i}nez-Gonz\'alez,$^1$
P. Vielva, $^3$ 
and L. Cay{\'o}n$^4$\\
$^1$IFCA, CSIC-Univ. de Cantabria, Avda. los Castros, s/n, 39005-Santander,
Spain\\
$^2$Dpto. de F\'{\i}sica Moderna, Univ. de Cantabria, Avda. los Castros, 
s/n, 39005-Santander, Spain\\
$^3$Physique Corpusculaire et Cosmologie, Coll\`ege de France, 11 pl. M. 
Berthelot, F-75231 Paris Cedex 5, France\\
$^4$Department of Physics, Purdue University, 525 Northwestern Avenue,
West Lafayette, IN 47907-2036, USA}

\date{Accepted  Received ; in original form }
\begin{document}
\maketitle

\begin{abstract}
An extremely cold and big spot in the WMAP 1-year data is analyzed. 
Our work is a continuation of a previous paper (Vielva et al. 2004) 
where non-Gaussianity was detected, with a method based on the 
Spherical Mexican Hat Wavelet (SMHW) technique. 
We study the spots at different thresholds on the SMHW coefficient maps, 
considering six estimators, namely number of maxima, 
number of minima, number of hot and cold spots, and number of pixels of the spots.
At SMHW scales around $4^\circ$ ($10^\circ$ on the sky), 
the data deviate from Gaussianity.
The analysis is performed on all sky, the northern and southern hemispheres, and on
four regions covering all the sky. A cold spot at ($b = -57^\circ, l = 209^\circ$) is found to be the
source of this non-Gaussian signature. We compare the spots of our data with
10000 Gaussian simulations, and conclude that only around 0.2\% of them present 
such a cold spot. Excluding this spot, the remaining map is compatible with Gaussianity
and even the excess of kurtosis in Vielva et al. 2004, is found to be due exclusively to this spot.
Finally, we study whether the
spot causing the observed deviation from Gaussianity
could be generated by systematics or foregrounds.
None of them seem to be responsible for the non-Gaussian detection.
\end{abstract}

\begin{keywords}
methods: data analysis - cosmic microwave background
\end{keywords}

\section{Introduction}

The Cosmic Microwave Background (CMB) is the most ancient image of the universe. The measured
temperature fluctuations reveal important characteristics of the primitive universe. Nowadays
many models try to explain the primordial density fluctuations, and the key for deciding 
among them is the Gaussianity of the temperature fluctuations of the CMB. According 
to standard inflationary theories, these fluctuations are generated by a Gaussian, homogeneous
and isotropic random field, whereas non-standard inflation (Linde \& Mukhanov 1997, Peebles 
1997, Bernardeau \& Uzan 2002 and Acquaviva et al. 2003) and topological defect 
models (Turok \& Spergel 1990 and Durrer 1999), predict non-Gaussian random fields.

Many difficulties have to be avoided in the Gaussianity analysis, since the temperature 
fluctuations may come from different sources, which should be identified. 
The primary fluctuations were originated
on the last scattering surface (LSS), while secondary fluctuations were produced in the 
trip that CMB photons cover from the LSS to us. Some examples of the latter are the 
fluctuations caused by the reionization of the universe (Ostriker \& Vishniac 1986 and 
Aghanim et al. 1996), the Rees-Sciama effect (Rees \& 
Sciama 1968 and Mart{\'\i}nez--Gonz{\'a}lez \& Sanz 1990), or
gravitational lensing (Mart{\'\i}nez--Gonz{\'a}lez et al. 1997, Hu 2000 and Goldberg \&
Spergel 1999). 
In addition we have systematic effects and contaminating foregrounds, 
like the Galactic emissions or extragalactic
point sources, which have to be considered in the Gaussianity study. 

In the last years, many groups have published analysis of Gaussianity. The Wilkinson Microwave
Anisotropy Probe (WMAP) 1-year all-sky data (Bennett et al. 2003a), were found to be consistent
with Gaussianity by the WMAP science team (Komatsu et al. 2003). A measure of the phase 
correlations of temperature fluctuations (taking  several combinations of the 
bispectrum into account), and the Minkowski functionals, were used in the latter paper. 
Thereafter several other groups have found non-Gaussian
signatures or asymmetries in the WMAP data
(Park 2003, Eriksen et al. 2003, 2004a, Hansen et al. 2004a, b).  
Several methods were used including Minkowski functionals,
\emph{N}-point correlation functions, or local curvature methods. 
Also some other works such as Chiang et al. 2003 and Copi et al. 2003, detect deviations from Gaussianity, using the 
Internal Linear Combination (ILC) map (Bennett et al. 2003a) and other maps, derived from the WMAP data. 
(However, the use of the ILC map is not recommended by the WMAP team, because of their complex noise and 
foreground properties).

Our basic reference is Vielva et al. 2004, where a non-Gaussian 
detection in the WMAP 1-year data was reported.
Convolving the data with the SMHW, an excess of kurtosis was found at scales around $4^\circ$, 
involving a size on the sky close to $10^\circ$.
The deviation from Gaussianity presented an upper tail probability of 0.4\%. This deviation
was localized in the southern hemisphere and an extremely cold spot at 
$b = -57^\circ, l = 209^\circ$, 
was regarded as a possible source of the non-Gaussianity (we call it \emph{The Spot}). 

Mukherjee \& Wang 2004 have carried out an extrema analysis using the SMHW, finding an excess of cold 
pixels in all the sky at the same scales as Vielva et al. 2004.  
In the present paper, our aim is to localize the non-Gaussianity, specifying whether The Spot alone is the
origin of the detection, and to study all the spots of the data in order to quantify how atypical 
The Spot under a Gaussian model is.

Also recently, Larson \& Wandelt 2004 have carried out an extrema analysis in real space, finding 
non-Gaussian evidences. 

The present paper is organized as follows. In $\S$2 we describe the data and 
simulations used to study the Gaussianity. The wavelet technique and the
estimators are explained in $\S$3. The results of our work are presented
in $\S$4. In $\S$5 we discuss the possible sources of the non-Gaussian
detection and in $\S$6 we present the conclusions of our results.

\section{Data and simulations}

We have followed a similar procedure as in Vielva et al. 2004. Hence our data are
the 1-year WMAP data, available in the \emph{Legacy Archive for Microwave 
Background Data Analysis} (LAMBDA) web page \footnote{http://www.cmbdata.gsfc.nasa.gov/},
choosing for our analysis the combined Q-V-W map. The WMAP receivers, 
take their observations at 5 frequency bands, namely K-band (22.8 GHz, 1 receiver), 
Ka-Band (33.0 GHz, 1 receiver), Q-Band (40.7 GHz, 2 receivers), V-Band (60.8 
GHz, 2 receivers) and W-Band (93.5 GHz, 4 receivers). The combined map is a linear combination of the
receivers where CMB is the dominant signal, namely Q, V, W radiometers: 
\begin{equation}
\label{eq:combination}
T({\mathbf{x}}) = \sum_{j = 3}^{10} 
{T_j}({\mathbf{x}})~{w_j}({\mathbf{x}}),
\end{equation}
The Q-Band receivers are denoted with indices $j = 3, 4$, V-Band receivers with indices $j = 5, 6$,
whereas indices $j = 7, 8, 9, 10$ correspond to the four W-Band receivers.
Noise weights ${w_j}({\mathbf{x}})$ are defined
so that receivers with less noise have more weight:
\begin{eqnarray}
\label{eq:noise}
w_j({\mathbf{x}}) = \frac{\bar{w}_j({\mathbf{x}})}{\sum_{j =
3}^{10}{\bar{w}_j}({\mathbf{x}})},~~ &
\bar{w}_j({\mathbf{x}}) = \frac{{N_j}({\mathbf{x}})}{{{\sigma_0}_j}^{2}}
\end{eqnarray}
where ${{\sigma_0}_j}$ is the noise dispersion per observation and 
${N_j}({\mathbf{x}})$ is the number of observations performed by the
receiver $j$ at position ${\mathbf{x}}$ (see Bennett et al. 2003a).
Foregrounds are corrected for each receiver, using the foreground 
template described in Bennett et al. 2003b. The combined map increases 
the signal to noise ratio.

The Hierarchical, Equal Area and 
iso-Latitude Pixelization (HEALPix, G\'{o}rski et al. 1999) is used in all maps. The resolution
parameter of the original map, is $N_{side}=512$ but we have degraded it to $N_{side}=
256$, since very small scales are dominated by noise. Afterwards the \emph{Kp0} 
mask was applied in order to get rid of foreground or point source contaminated
pixels. This mask removes 23.2\% of the pixels, especially the strongly contaminated
pixels around the Galactic plane. Finally, the residual monopole and dipole outside the mask are
subtracted. 
The mask and the foreground cleaned frequency maps can be found in the LAMBDA site.

Following the same procedure as in Vielva et al. 2004, we have produced 10000 Gaussian simulations
with the purpose of checking the Gaussianity of the data. Starting with the cosmological parameters
estimated by the WMAP team (Table 1 of Spergel et al. 2003), we have calculated the $C_\ell$ 
using CMBFAST (Seljak \& Zaldarriaga 1996). We have generated random Gaussian $a_{\ell m}$ of CMB 
realizations and convolved them with the adequate beam for each receiver. 
After transforming from harmonic to
real space, uncorrelated Gaussian noise realizations have been added, taking into 
account the number of observations per pixel (${N_j}({\mathbf{x}})$) and the noise dispersion per 
observation (${{\sigma_0}_j}$). At this point the 
simulations were degraded to  $N_{side}=256$ and the \emph{Kp0} mask has been applied. 
At the end, monopole and dipole are subtracted outside the mask

In the present work we have studied the spots above or below a given threshold $\nu$, 
therefore we have normalized the data and the simulations, dividing each map by its 
dispersion, after subtracting the mean. 

\section{Wavelets and estimators}

Several works have been published in recent years concerning CMB studies based on wavelets.
Wavelet transforms can be applied to separate the different components appearing in
microwave observations (Tenorio et al. 1999, 
Cay{\'o}n et al. 2000, Vielva et al. 2001a, Vielva et al. 2001b and Vielva et al. 2003a), 
in denoising techniques (Sanz et al. 1999a,b) and in Gaussianity studies
(Pando et al 1998, Hobson et al. 1999, Aghanim et al. 2003, Barreiro et al. 2000, 
Barreiro \& Hobson 2001 and references listed below for the spherical symmetry). 
Wavelets increase the signal to noise ratio, allowing us to detect weak non-Gaussian signals. 
Furthermore they preserve the spatial location 
and the angular scale \emph{R} of a hypothetical non-Gaussian feature.
This is a very important advantage since our main goal in this paper is to study 
the hot and cold spots over several thresholds.

In our case, the ideal wavelet is the SMHW (Antoine J. P. \& Vanderheynst P. 1998), 
which has been applied to non-Gaussian studies
of the COBE-DMR data (Cay{\'o}n et al. 2001, 2003), to \emph{Planck} simulations 
(Mart{\'\i}nez-Gonz{\'a}lez et al. 2002) and to the 1-year WMAP data in our basic
reference (Vielva et al. 2004).
The SMHW optimally enhances the non-Gaussian signatures on the sphere
(Mart{\'\i}nez--Gonz{\'a}lez et al. 2002).
The expression of the SMHW is:
\begin{equation}
\label{eqSMHW}
   \Psi_S(y,R) = 
\frac{1}{\sqrt{2\pi}N(R)}{\Big[1+{\big(\frac{y}{2}\big)}^2\Big]}^2
  \Big[2 - {\big(\frac{y}{R}\big)}^2\Big]e^{-{y}^2/2R^2},
\end{equation}
where $R$ is the scale and $N(R)$ is a normalization constant: 
$N(R)\equiv R\sqrt{1 + R^2/2 + R^4/4}$.
The distance $y$ on the tangent plane is related to the polar 
angle ($\theta$) as: $y\equiv 2\tan{\theta/2}$.
Data and simulations have been convolved with the SMHW at different scales, to obtain the
SMHW coefficient maps analyzed in this work.   
The study has also been performed in the data
and simulations previous to the convolution with the SMHW. These
maps are referred to as maps in real space, at scale $R_o$.

We have considered the following six estimators: number of maxima, number of minima, number
of hot spots, number of cold spots, number of hot pixels 
(hot area) and number of cold pixels (cold area).
All of them are referred to a particular threshold; thus maxima, hot spots and hot pixels
lie above a threshold $\nu$, whereas number of minima, cold spots and cold pixels 
lie below $-\nu$.
The analysis has been done in seven regions: all sky, northern hemisphere, southern
hemisphere, Northeast $(b>0,l<180)$, Northwest $(b>0,l>180)$, Southeast $(b<0,l<180)$
and Southwest $(b<0,l>180)$. Each region has its own dispersion and has to be 
normalized separately.

\section{The analysis}

The first step in our analysis is the convolution of data and simulations with the
SMHW at 15 scales ($R_1 = 13.7$, $R_2 = 25$, $R_3 = 50$, $R_4 = 75$, 
$R_5 = 100$, $R_6 = 150$, $R_7 = 200$, $R_8 = 250$, $R_9 = 300$, $R_{10} = 400$, 
$R_{11} = 500$, $R_{12} = 600$, $R_{13} = 750$, $R_{14} = 900$ and $R_{15} = 1050$ 
arcmin). 
Since the convolution mixes up pixels of the masked region with 
unmasked pixels, the resulting map has many corrupted SMHW coefficients. This effect involves
a loss of efficiency, so we applied an additional mask as explained in Vielva et al. 2004,
extending the \emph{Kp0} mask to 2.5 \emph{R}. The results in Vielva et al. 2004 were shown
to be quite independent of the definition of this extended mask. For each map we have calculated the
estimators on the seven previously described regions. The chosen thresholds are $\pm 2.0$,
$\pm 2.5$, $\pm3.0$, $\pm 3.5$, $\pm 4.0$ and $\pm 4.5$. In the figures we have represented only
absolute values for the thresholds, recalling that for cold spots and minima, values lie below a 
negative threshold, whereas hot spots and maxima values lie above a positive threshold.
 
Once the estimators were calculated for data and simulations, we proceeded to 
establish acceptance intervals at significance levels $\alpha$ (32\%,5\% and 1\%).  
Therefore we have sorted the $10000$ values of each estimator into ascending numbers
and excluded $10000\alpha/2$ values from each tail of the distribution. The two limiting values,
corresponding to simulations $1+10000\alpha/2$ and $10000 (1-\alpha/2)$
define the acceptance interval containing a probability of $1-\alpha$. The remaining probability
is $\alpha/2$ above and below the interval. 
In all figures this acceptance intervals are
plotted, the 32\% interval corresponds to the inner band, the 5\% interval to the middle band and the 1\% 
significance level, to the outer one. 

The results for the different regions are presented in the next three subsections.

\subsection{All sky, North and South}
Deviations from Gaussianity have been detected at scales $R_{5}$, $R_{8}$ and $R_{9}$.
\begin{figure*}
\begin{center}
\includegraphics[width=8cm]{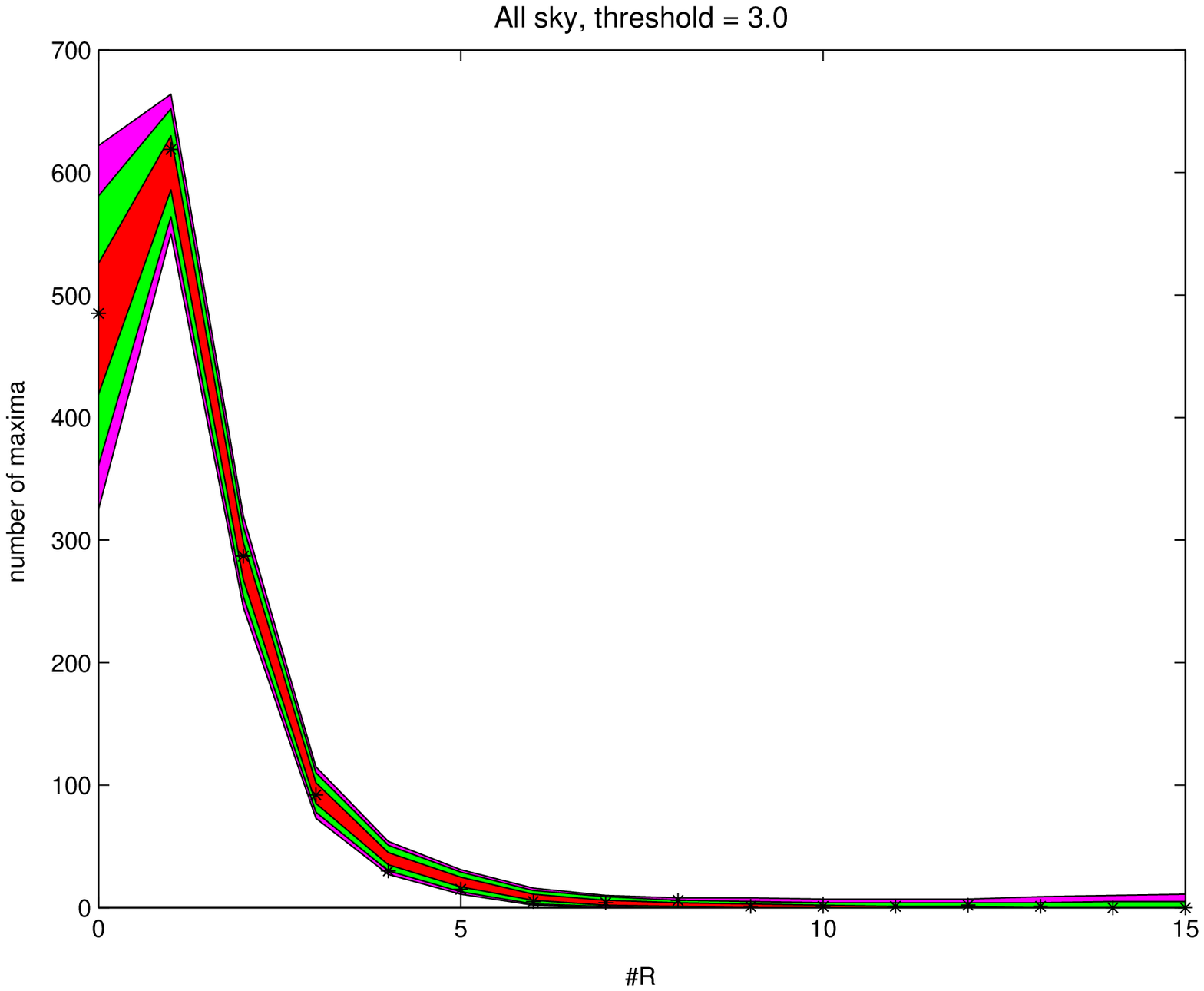}
\includegraphics[width=8cm]{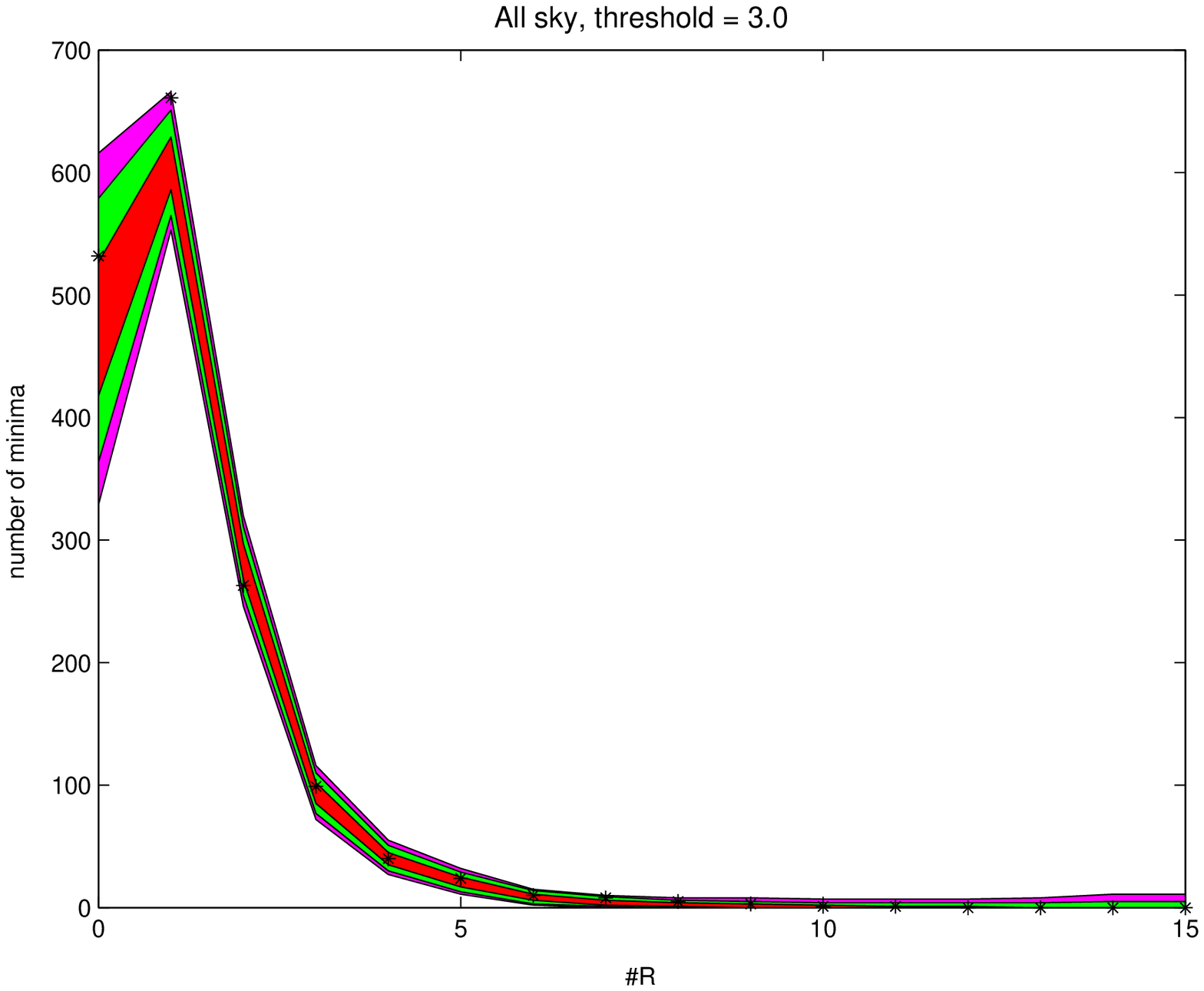}
\includegraphics[width=8cm]{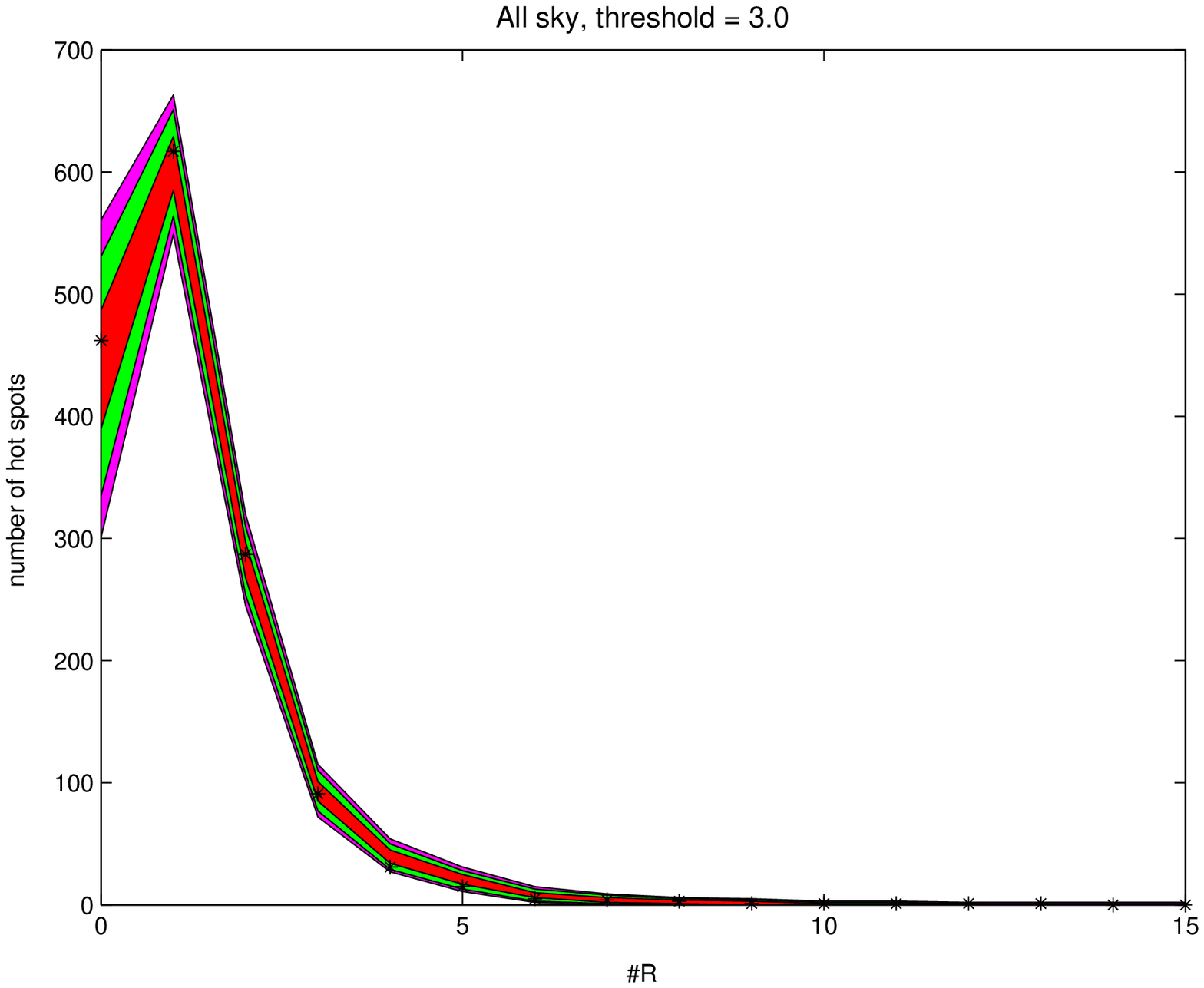}
\includegraphics[width=8cm]{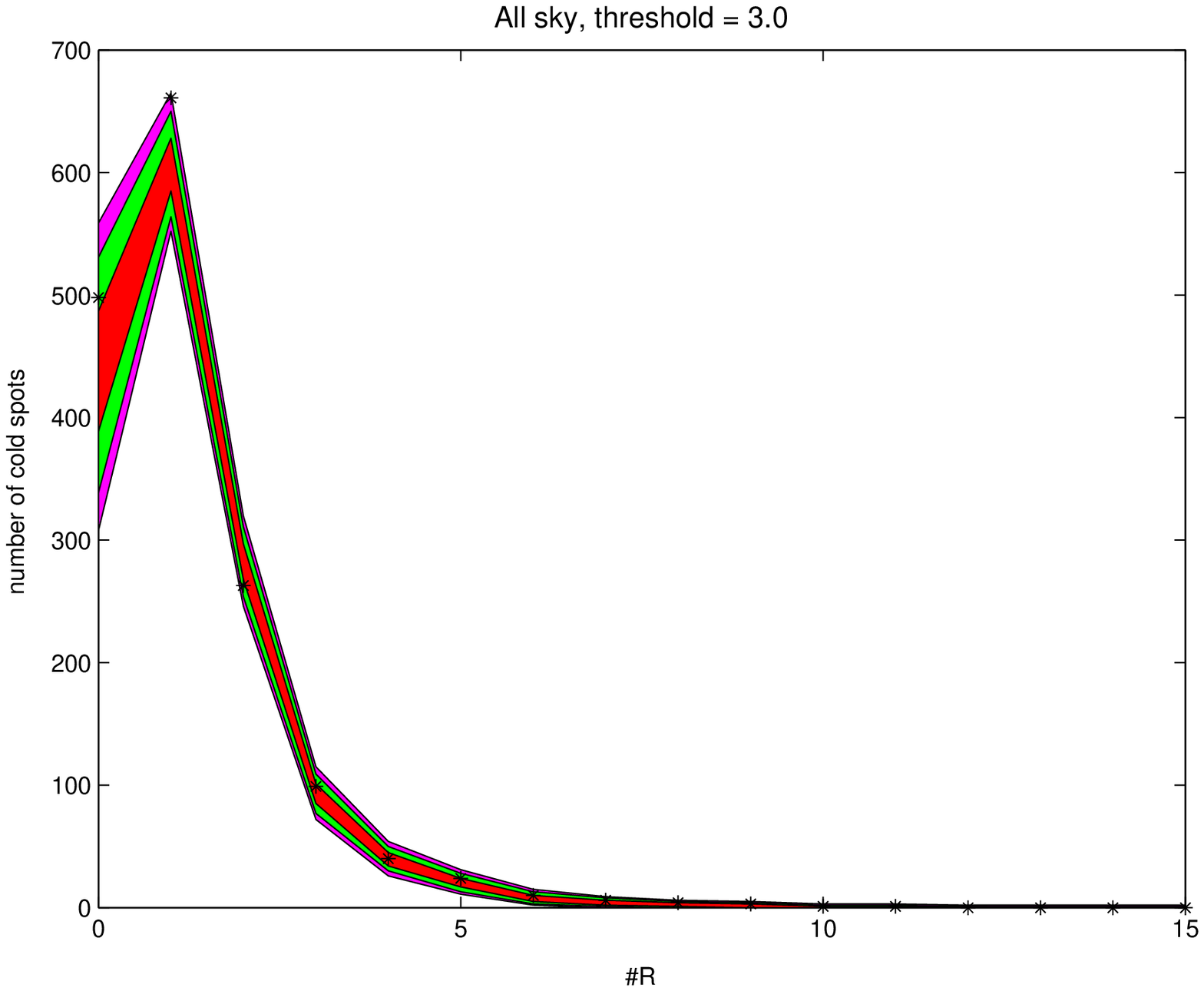}
\includegraphics[width=8cm]{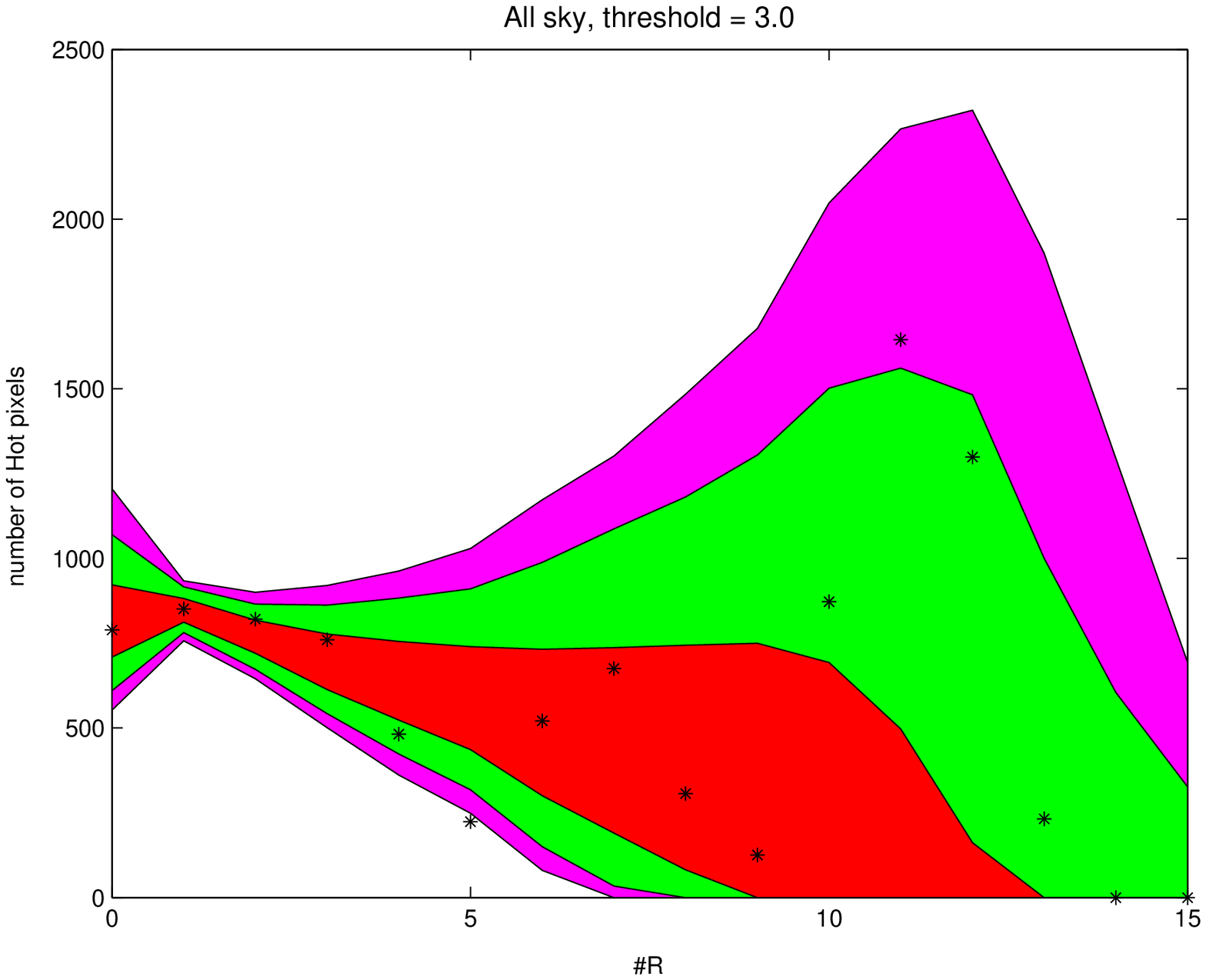}
\includegraphics[width=8cm]{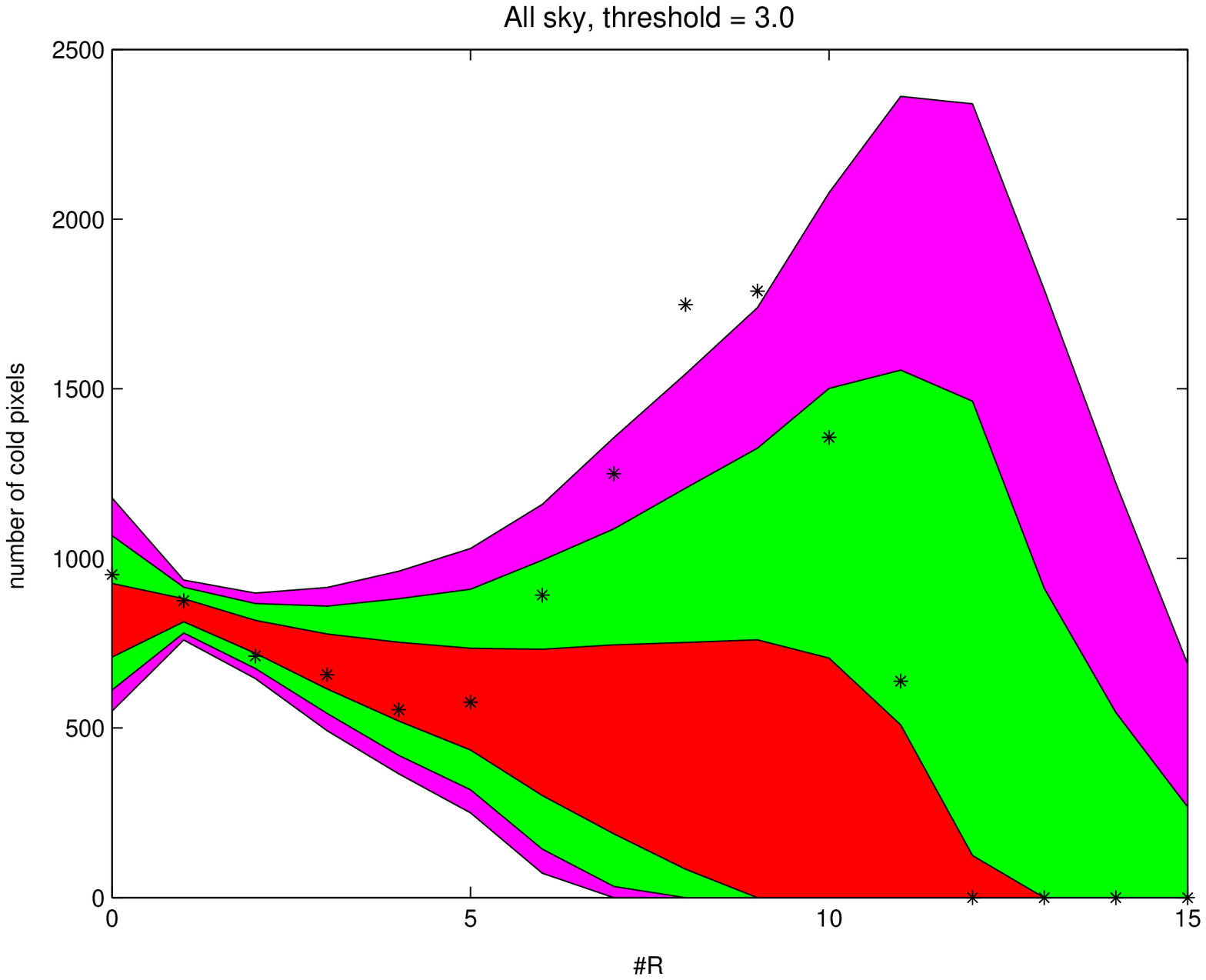}
\end{center}
\caption{This figure shows from left to right and top to bottom, number of maxima,
number of minima, number of hot spots, 
number of cold spots, hot and cold area for each of the considered scales at threshold 3.0. 
Scale $R_{0}$ corresponds 
to real space. The combined map values are plotted as asterisks. The acceptance intervals
for the 32\% (inner), 5\% (middle) and 1\% (outer) significance levels,
given by the 10000 simulations are also plotted.
Non-Gaussianity is found at scales $R_{5}$ in the hot area, $R_{8}$ and $R_{9}$ 
in the cold area. At other scales the data always lie at least within the 99\% acceptance band. 
For scale $R_{5}$ the deviation occurs only for this threshold and scale whereas scales 
$R_{8}$ and $R_{9}$ show a more significant detection.}
\label{fig:all_scales}
\end{figure*}
For threshold 3.0, the results are represented in Figure ~\ref{fig:all_scales}.
At scale $R_{5}$ the hot area lies outside the 99\% acceptance interval. However we have 
studied this case carefully, noting that this threshold is the only one were data lie 
outside the intervals, whereas
the number of hot spots and number of maxima do not depart from Gaussianity, even at this scale.
Dividing the sky into two hemispheres we checked that this was not a localized effect either. 
Even contiguous scales were compatible with Gaussianity, hence we concluded that this non-Gaussian
feature was not significant.

On the contrary, in the case of cold areas, deviations are
observed at scales $R_{8}$ and $R_{9}$ at several thresholds as can be
seen in Figure 2.
\begin{figure*}
\begin{center}
\includegraphics[width=8cm]{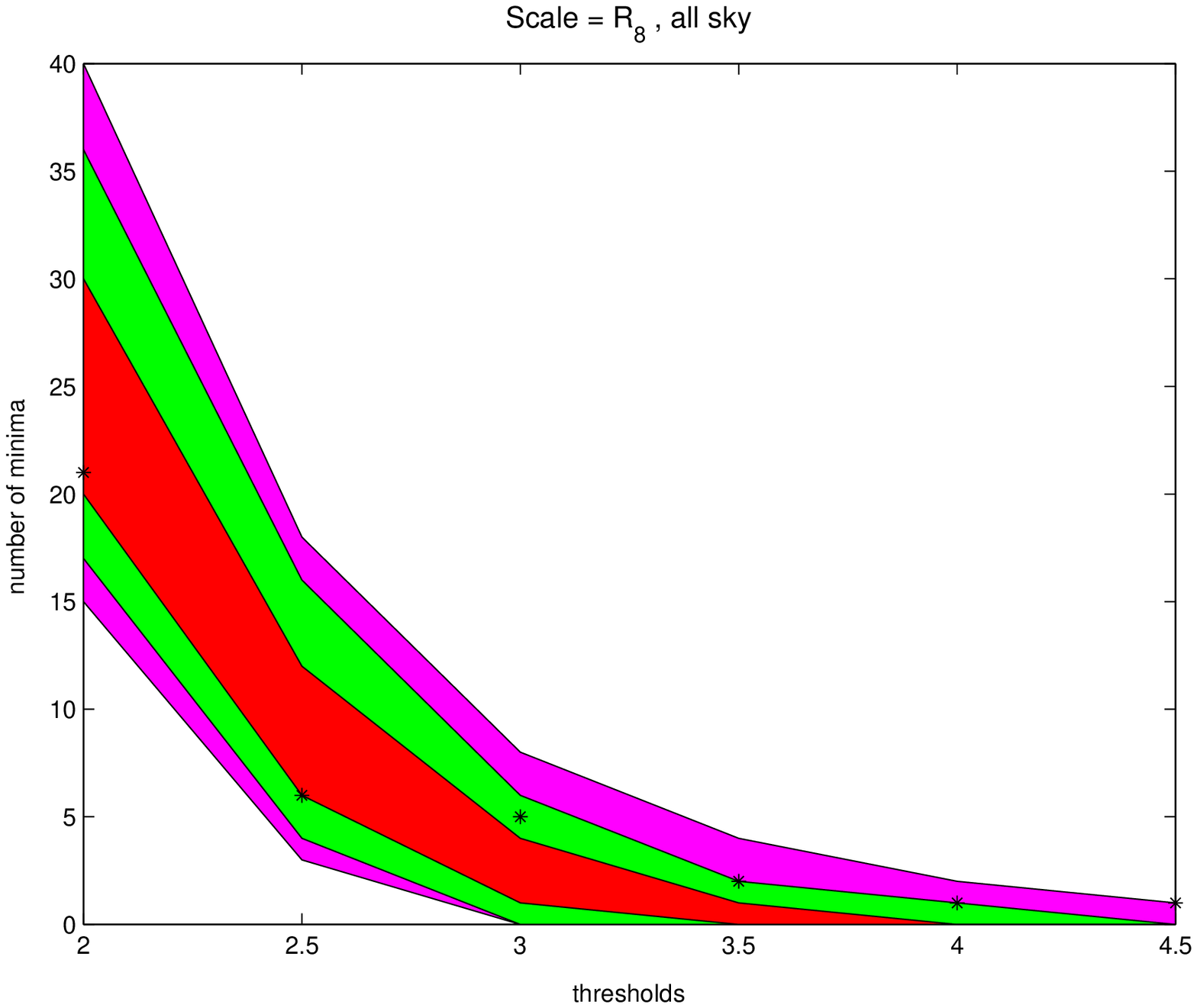}
\includegraphics[width=8cm]{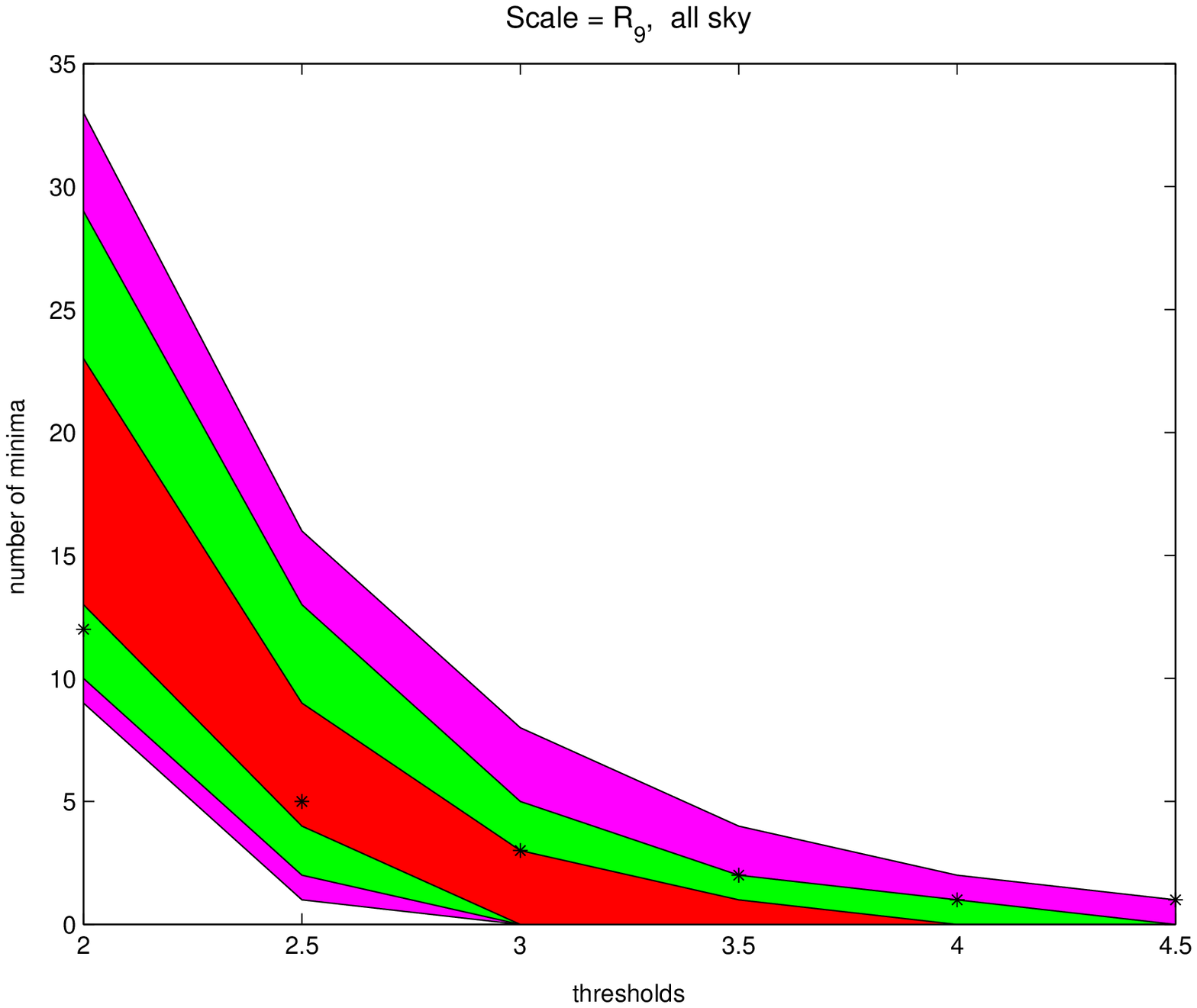}
\includegraphics[width=8cm]{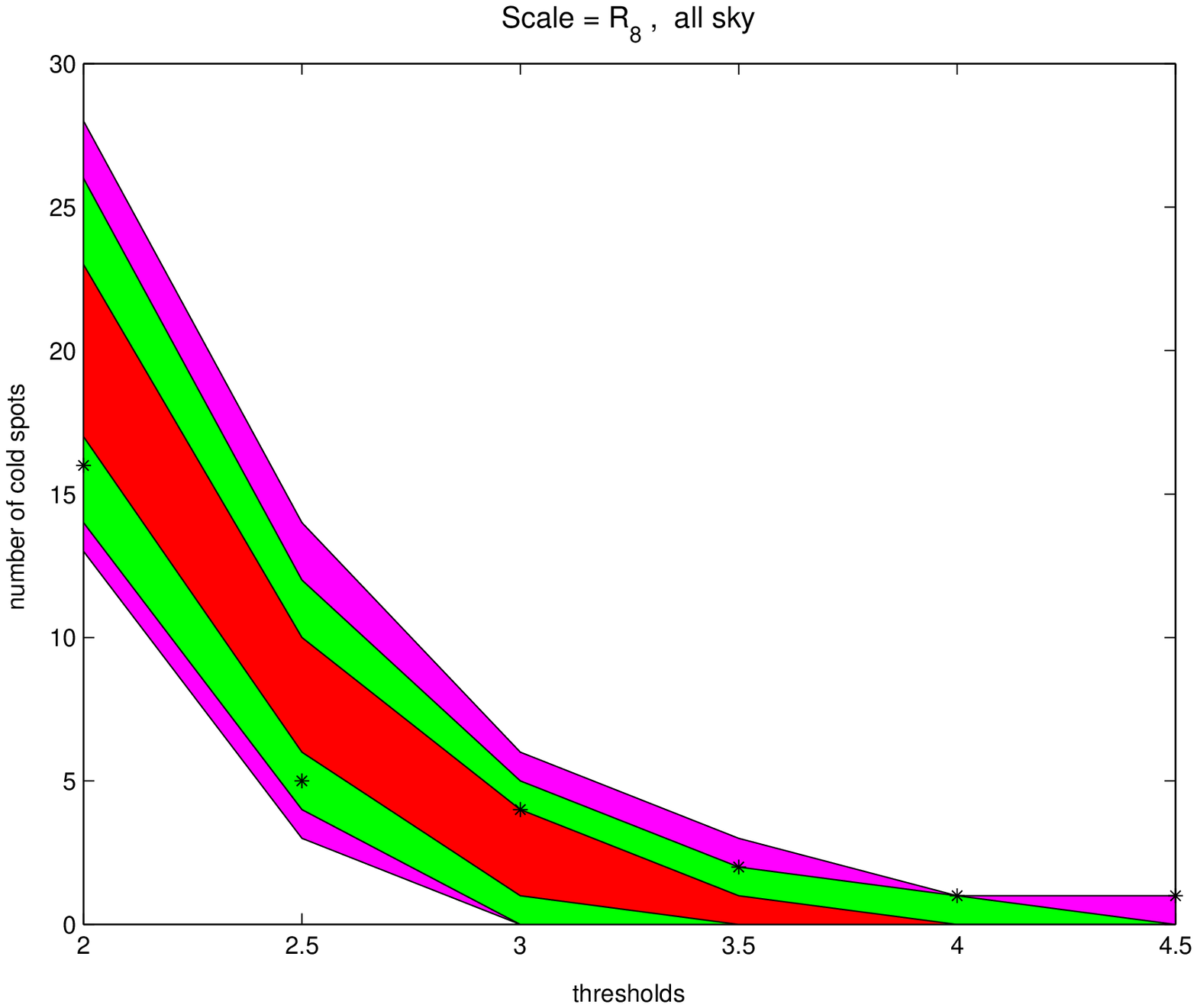}
\includegraphics[width=8cm]{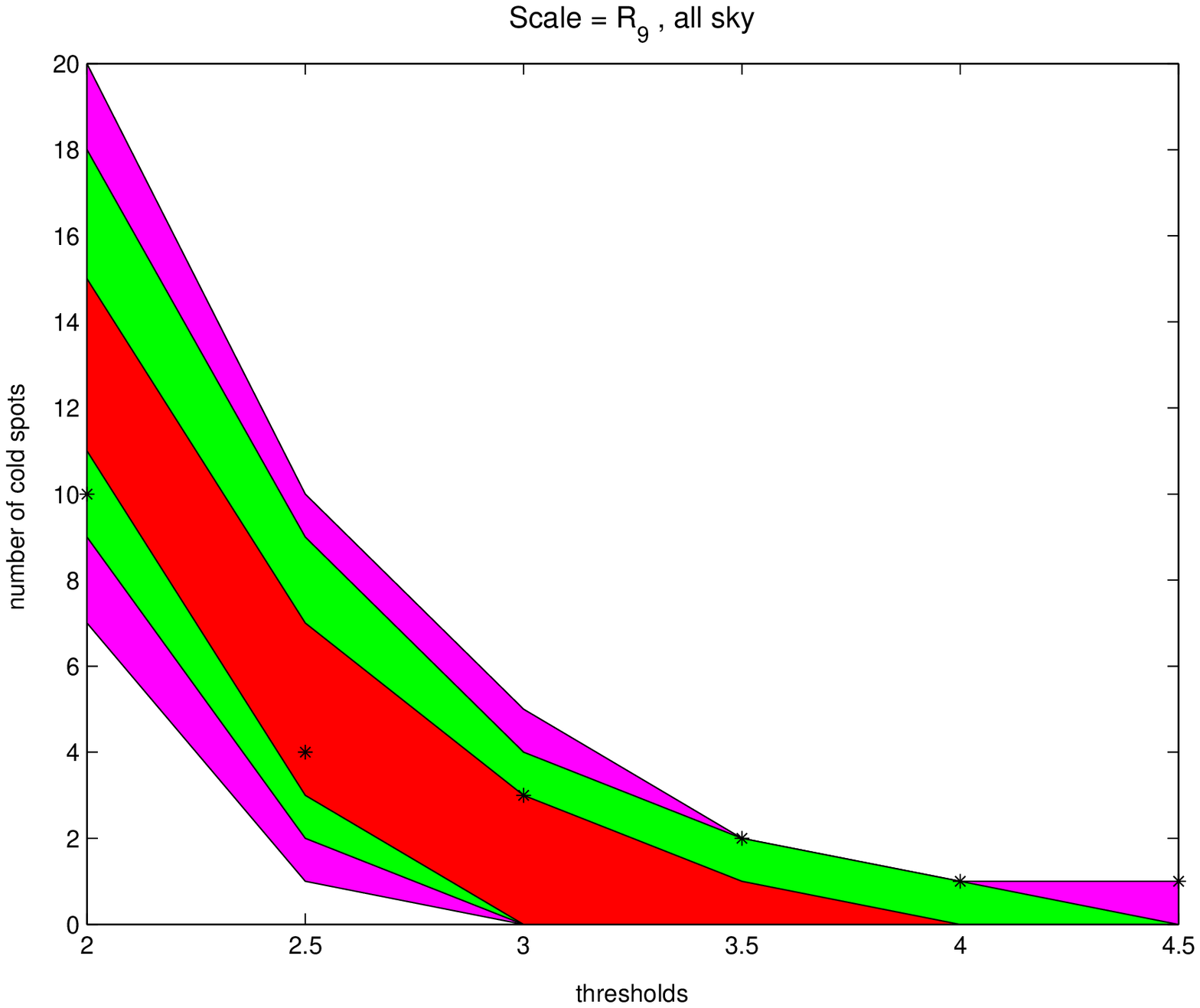}
\includegraphics[width=8cm]{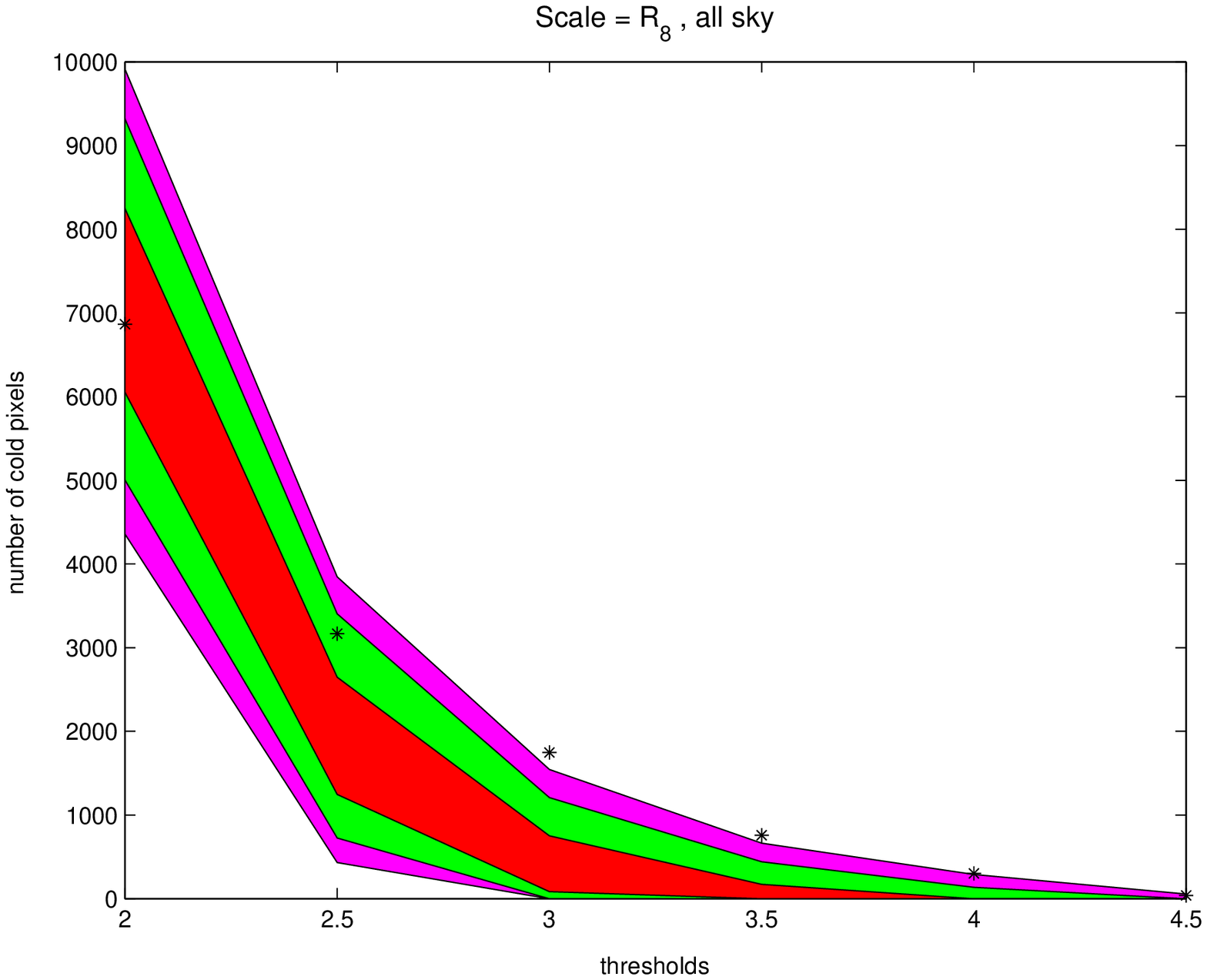}
\includegraphics[width=8cm]{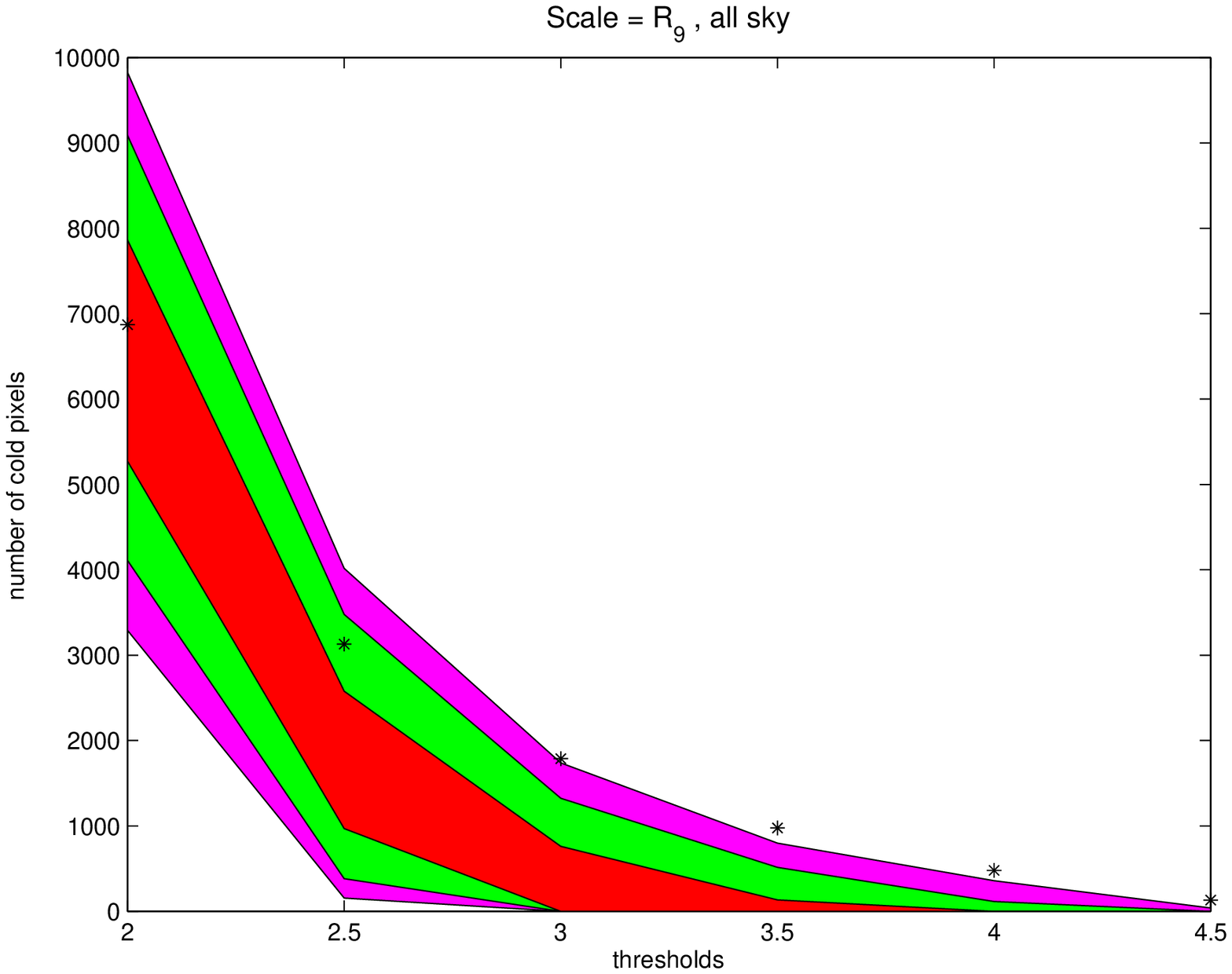}
\end{center}
\caption{Number of minima, number of cold spots and cold area are shown for scale $R_{8}$ at
the left column and scale $R_{9}$ at the right column. As in figure ~\ref{fig:all_scales} the
asterisks represent the combined WMAP data, and the acceptance intervals
for the 32\% (inner), 5\% (middle) and 1\% (outer) significance levels,
given by the 10000 simulations are also shown. 
The data lie outside the acceptance intervals for the area, at thresholds over 3.0. The borderline
of the  5\% and 1\% acceptance intervals coincide at some high thresholds for the number of spots,
because at this scales the number of spots is very low.}
\label{fig:scales_08_09}
\end{figure*}
We represent in Figure ~\ref{fig:scales_08_09} the number of cold spots, 
cold area and number of minima, for all considered thresholds at  
scales $R_{8}$ and $R_{9}$, in order to observe what is happening there with more detail. 
The number of minima and cold spots was exactly 
one at thresholds 4.0 and 4.5, reaching the borderline of our 99\% acceptance interval.
For the number of cold spots, at threshold 4.0 for scale $R_{8}$ and thresholds 3.5 and
4.0 for scale $R_{9}$, the 95\% and 99\% acceptance intervals coincide. This can happen when 
the number of spots is very low. Consider for example a binomial distribution where the 
number of spots can only take values 0 or 1. The way we define the acceptance intervals
determines that for such cases some of these intervals must coincide. 
The most striking non-Gaussian signature was 
found for the cold area at scales $R_{8}$ and $R_{9}$ where the data
exhibited an extremely high number of cold pixels at thresholds over 3. 
The cold area lies outside the 99\% acceptance interval at thresholds above 3.0 in the two
scales presented in Figure ~\ref{fig:scales_08_09}.

At thresholds 4.0 and 4.5 the mentioned observations reveal that the non-Gaussianity is only 
due to one particular spot, which reaches a
minimum value of $-4.7\sigma$ at ($b = -57^\circ, l = 209^\circ$) and scale $R_{9}$. 
At lower thresholds several spots contribute to the observed deviation. 
A precise analysis of The Spot is presented in the following sections. The data suggest that
we are dealing with a very cold and big spot.
These results agree with the results reported in Vielva et al. 2004, since the 
non-Gaussianity has been detected at the same scales.
\begin{figure*}
\begin{center}
\includegraphics[width=8cm]{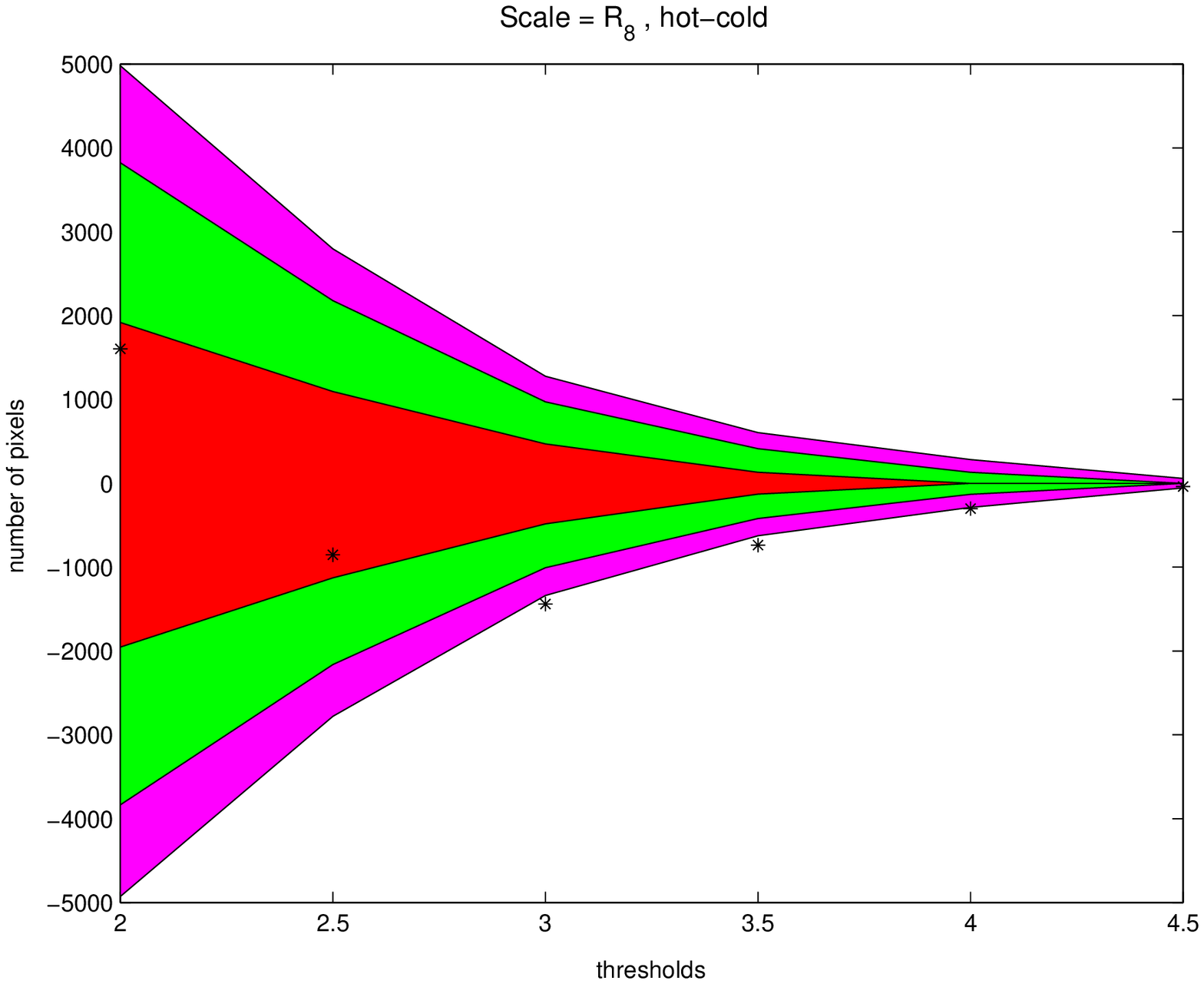}
\includegraphics[width=8cm]{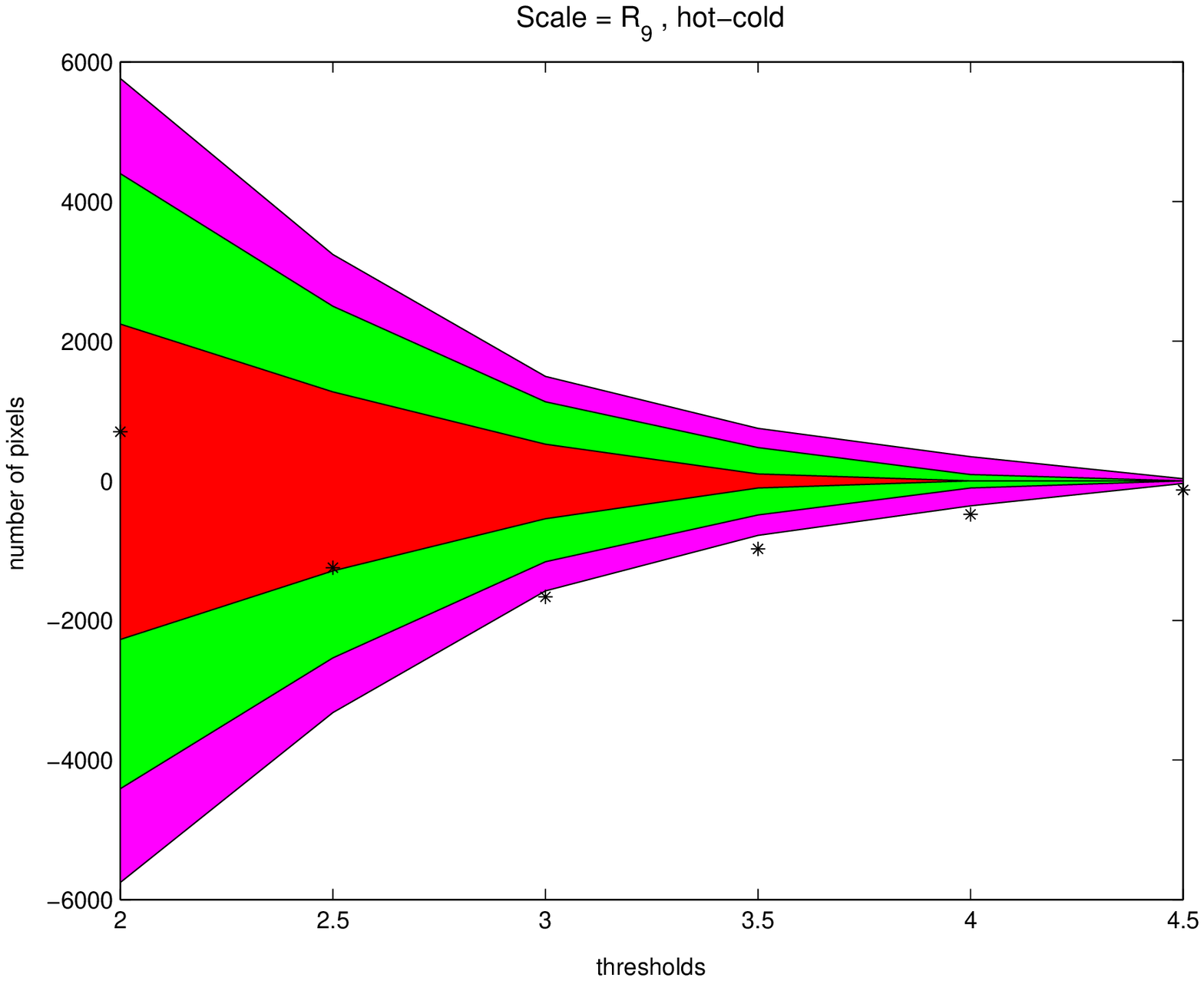}
\includegraphics[width=8cm]{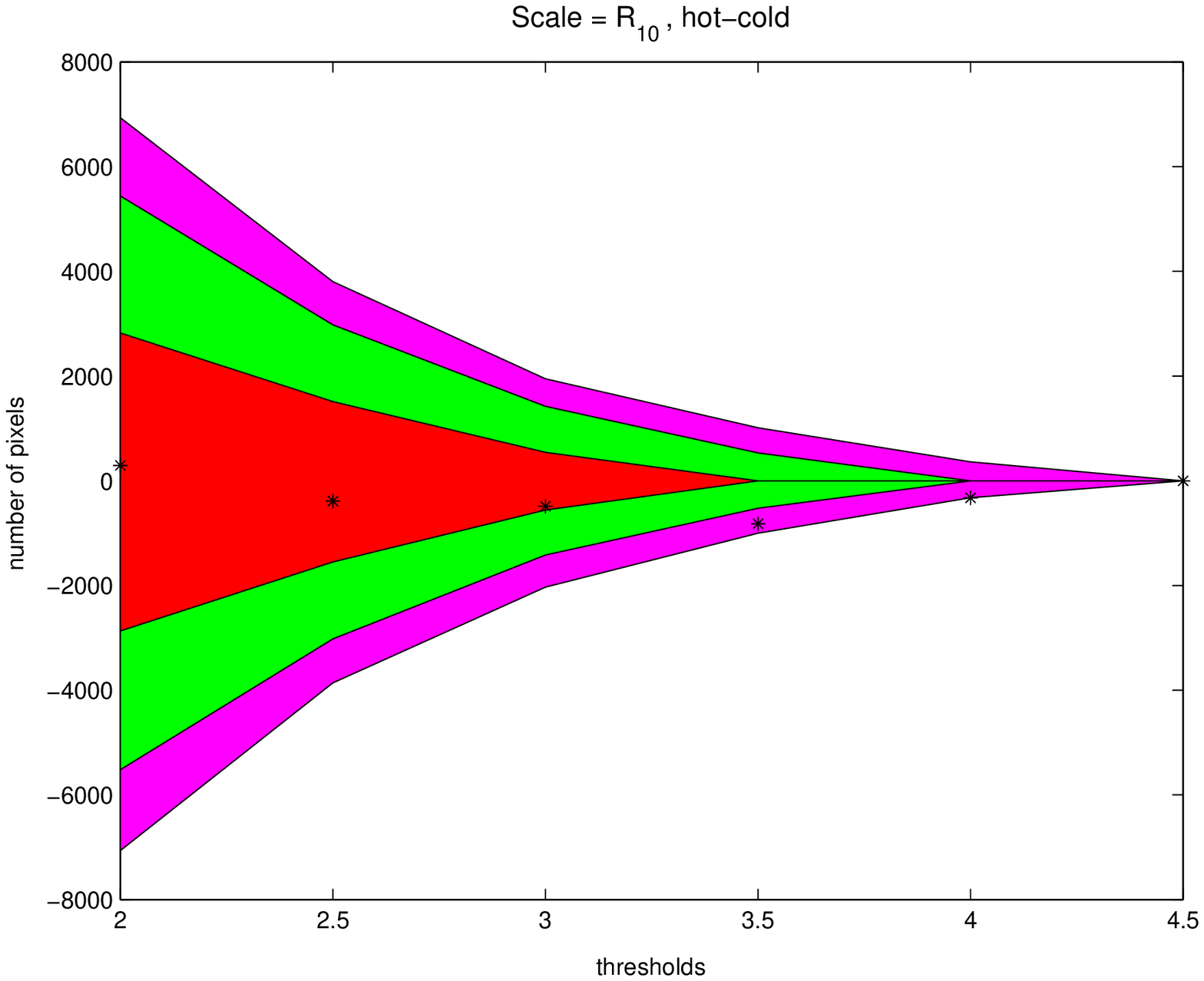}
\end{center}
\caption{Hot-cold asymmetry, in the area. The excess of cold pixels at scales $R_{8}$, 
$R_{9}$ and $R_{10}$ happens at high thresholds. Data and acceptance intervals are 
represented in the same manner as in previous figures.}
\label{fig:Asim_HC}
\end{figure*}
The hot spots did not show any non-Gaussian evidence. Furthermore, subtracting the cold pixels
from the hot pixels, a strong hot-cold asymmetry was revealed at scales $R_{8}$, $R_{9}$ and
$R_{10}$, see Figure ~\ref{fig:Asim_HC}.
\begin{table}
   \begin{center}
         \begin{tabular}{|c|c|c|}
	 \hline
	 Scale & threshold & probability \\
	 $R_{8}$ & 3.0 & 0.37\% \\
	         & 3.5 & 0.26\% \\
	         & 4.0 & 0.44\% \\
	         & 4.5 & 0.65\% \\
	 \hline
	 $R_{9}$ & 3.0 & 0.39\% \\
	         & 3.5 & 0.15\% \\
	         & 4.0 & 0.19\% \\
	         & 4.5 & 0.22\% \\
	 \hline
         $R_{10}$ & 3.0 & 18.22\% \\
	          & 3.5 & 1.04\% \\
	          & 4.0 & 0.48\% \\
	 \hline
      \end{tabular}
      \caption{Lower tail probabilities of having the hot-cold asymmetry (in the number of pixels) of 
               our data under a Gaussian model,
               at different scales and thresholds. Most of the values are below 1\%.
	       At scale $R_{10}$ and threshold 4.5
               the number of hot and cold pixels are both zero, because the threshold is
               too high and therefore we do not show this threshold for the mentioned scale.}
      \label{table:AsimHC}
    \end{center}
\end{table}
The lower tail probabilities displayed in Table ~\ref{table:AsimHC} show non-Gaussian values of
up to 99.85\%.
  
Our purpose was to locate the non-Gaussian sources. We studied the 
northern and southern hemispheres separately, expecting to find non-Gaussian results in the 
southern hemisphere because The Spot is located there. 
\begin{figure*}
\begin{center}
\includegraphics[width=10cm]{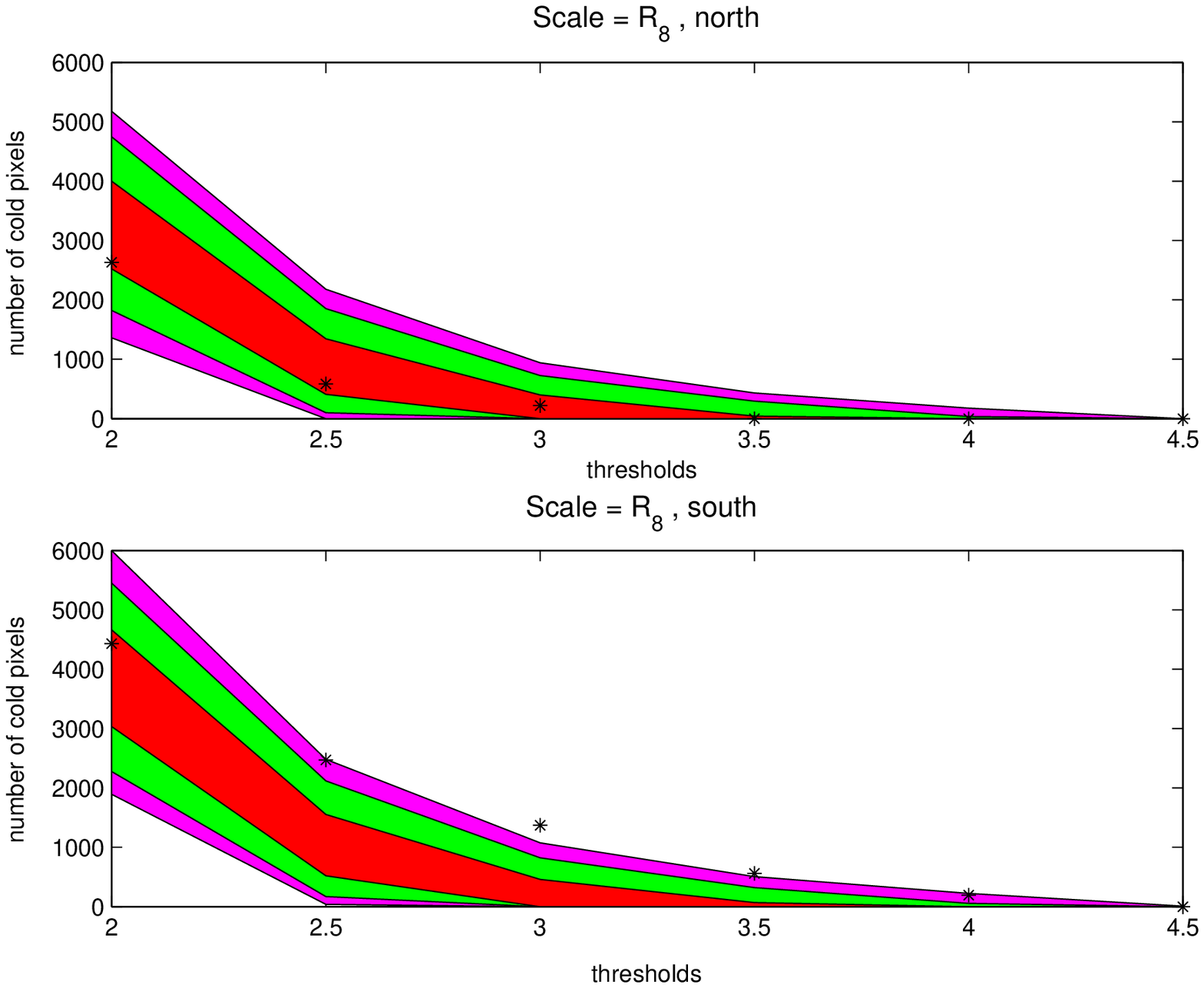}
\includegraphics[width=10cm]{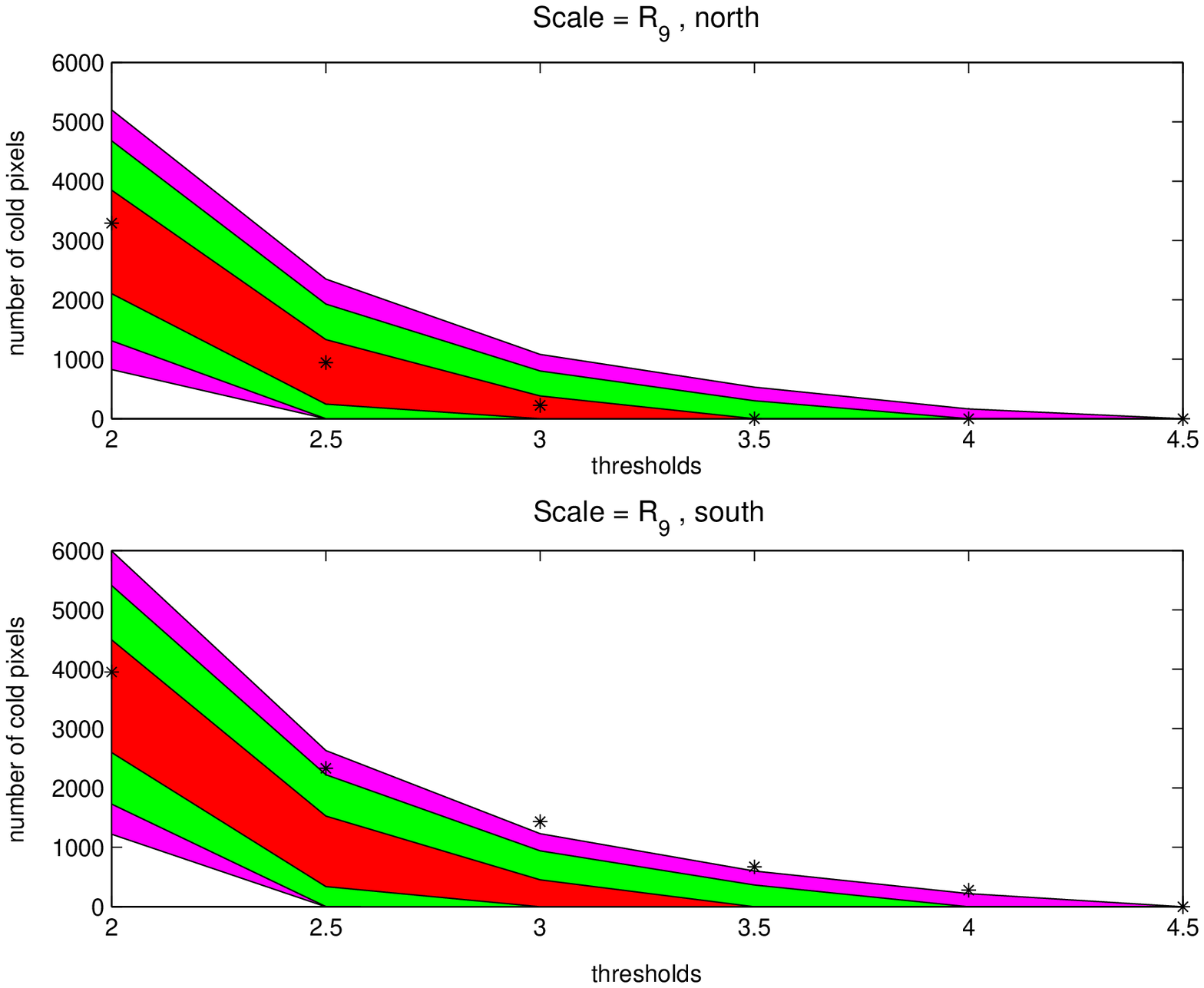}
\end{center}
\caption{Cold area in the northern and southern hemispheres
at scales $R_{8}$ and $R_{9}$. The southern hemisphere shows
a non-Gaussian behaviour. Again we represent data and acceptance intervals as in 
Figure ~\ref{fig:all_scales}.}
\label{fig:Area_Cold2}
\end{figure*}
Results presented in Figure ~\ref{fig:Area_Cold2} show that the northern
hemisphere is compatible with the Gaussian simulations whereas the southern hemisphere
shows a clear deviation from the acceptance intervals.
\begin{figure*}
\begin{center}
\includegraphics[width=10cm]{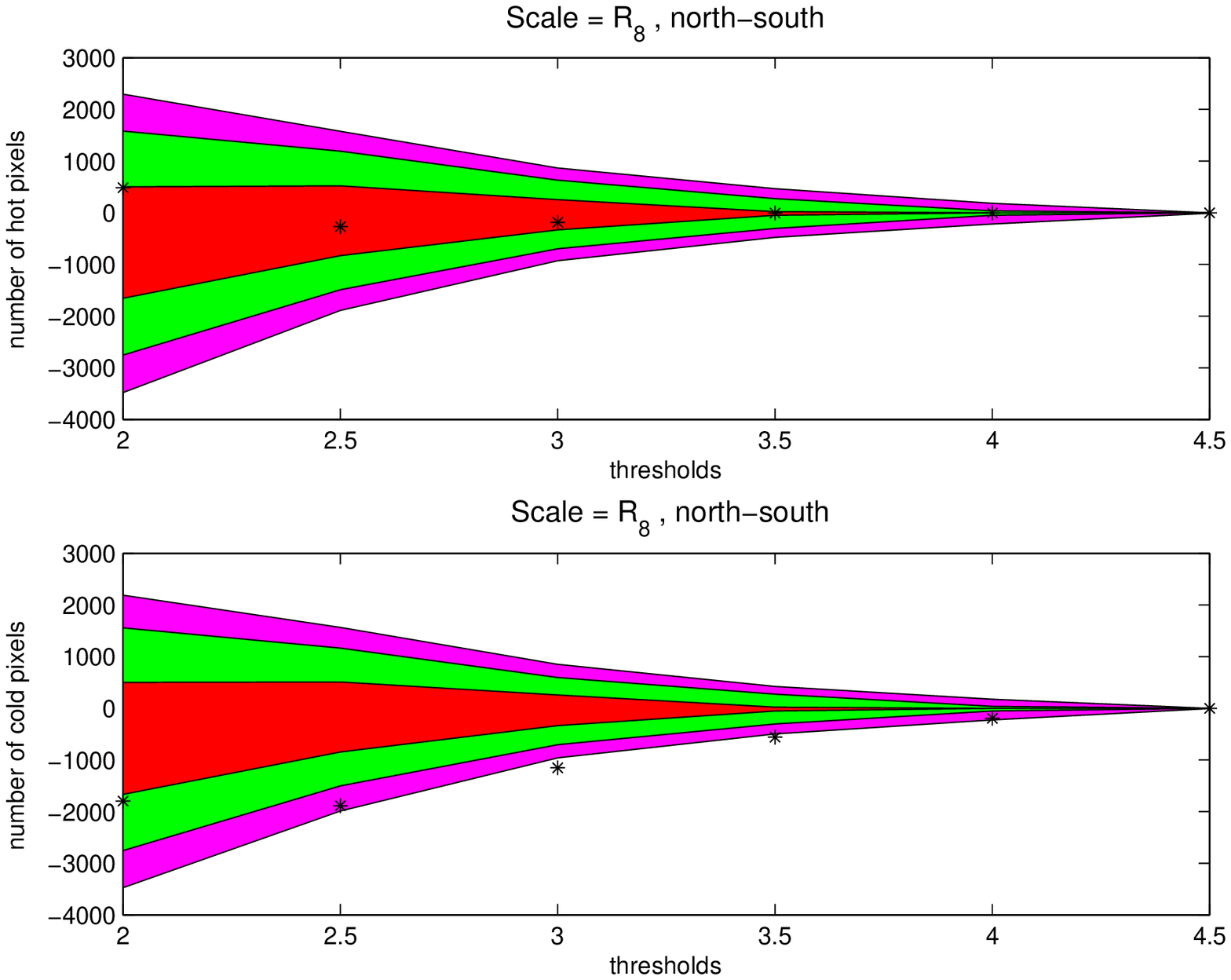}
\includegraphics[width=10cm]{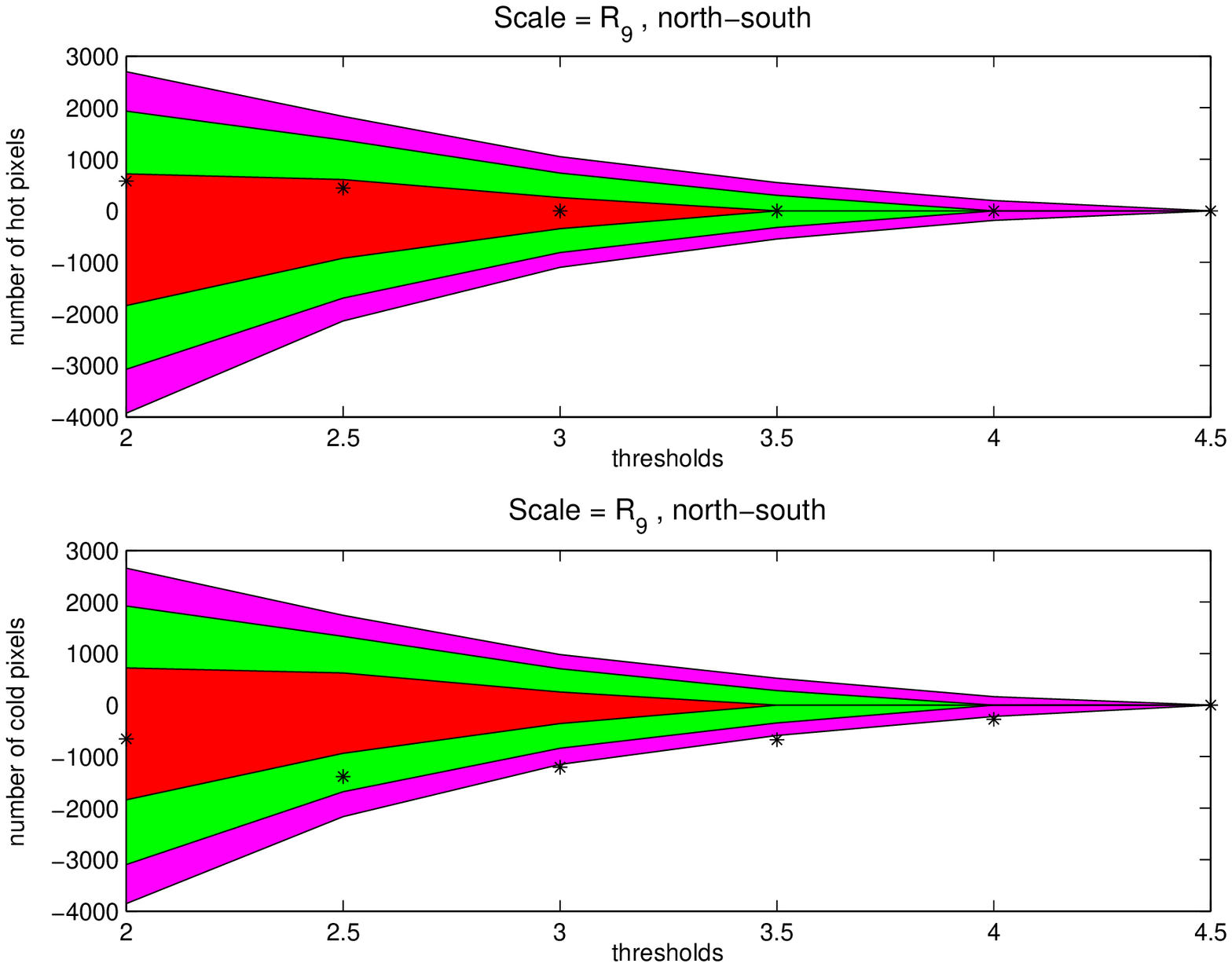}
\end{center}
\caption{The two upper panels represent the North-South asymmetry for hot and cold area, at scale $R_{8}$ 
and the two pannels below, the mentioned asymmetries at scale $R_{9}$. The southern hemisphere
shows an excess of cold pixels. Acceptance intervals and data are represented as in previous 
figures.}
\label{fig:Asim_Area}
\end{figure*}
To show the North-South asymmetry, the number of hot and cold pixels in the South
were subtracted from the Northern ones. Results are presented in Figure ~\ref{fig:Asim_Area}.
Deviations from Gaussianity can be clearly observed in the number of
cold pixels at thresholds 3.0 and above.
Previous works such as Vielva et al. 2004, Eriksen et al. 2004a or Hansen 2004b have already 
reported North-South asymmetries, which are confirmed here.
 
The marked observed asymmetry, prompted us to divide the sky into four regions with the purpose
of studying a Southeast-Southwest asymmetry.

\subsection{Four regions}
We analyzed four regions independently by splitting each hemisphere 
into two parts, $l<180$ (East)  and $l>180$ (West). 
\begin{figure*}
\begin{center}
\includegraphics[width=10cm]{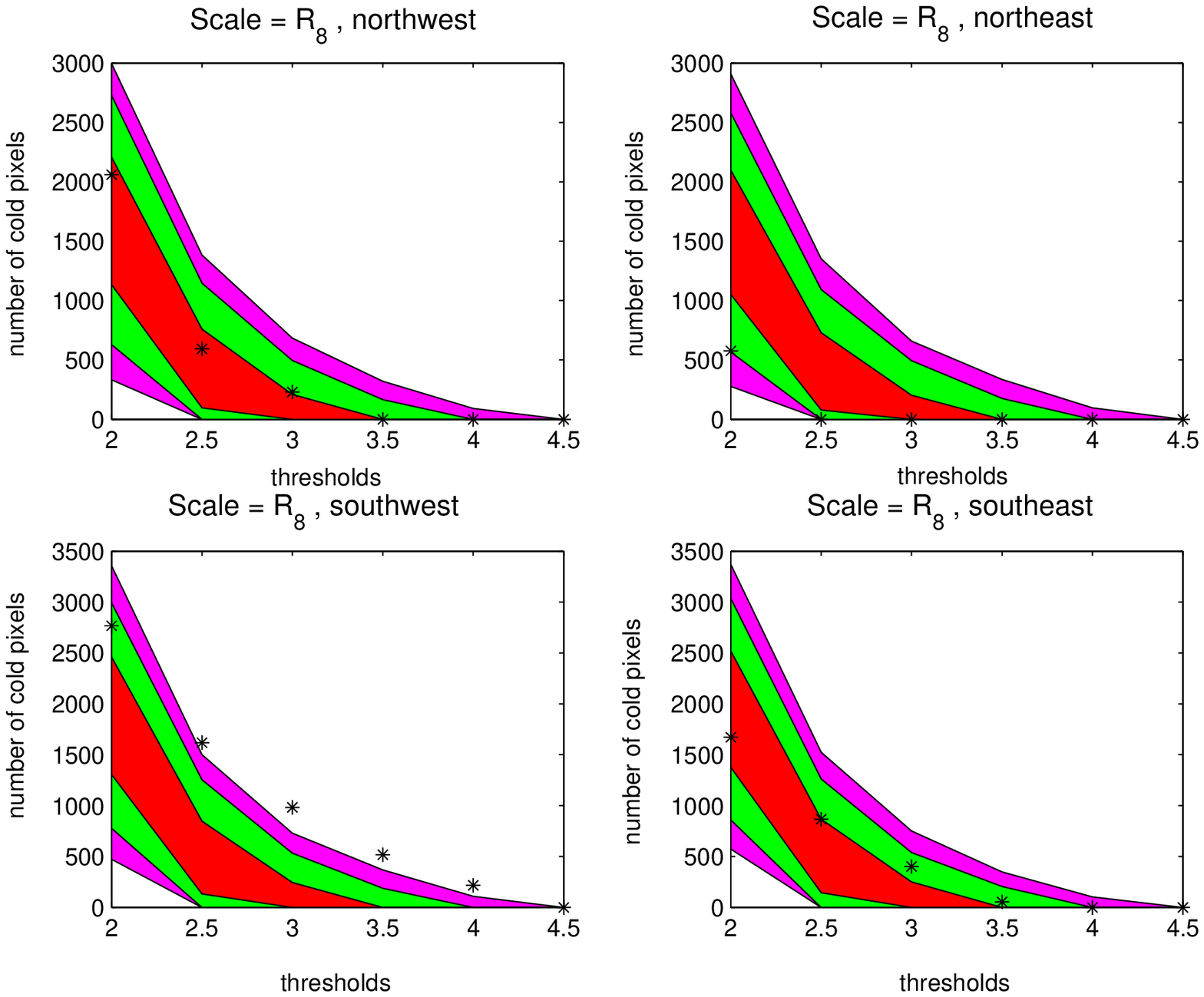}
\includegraphics[width=10cm]{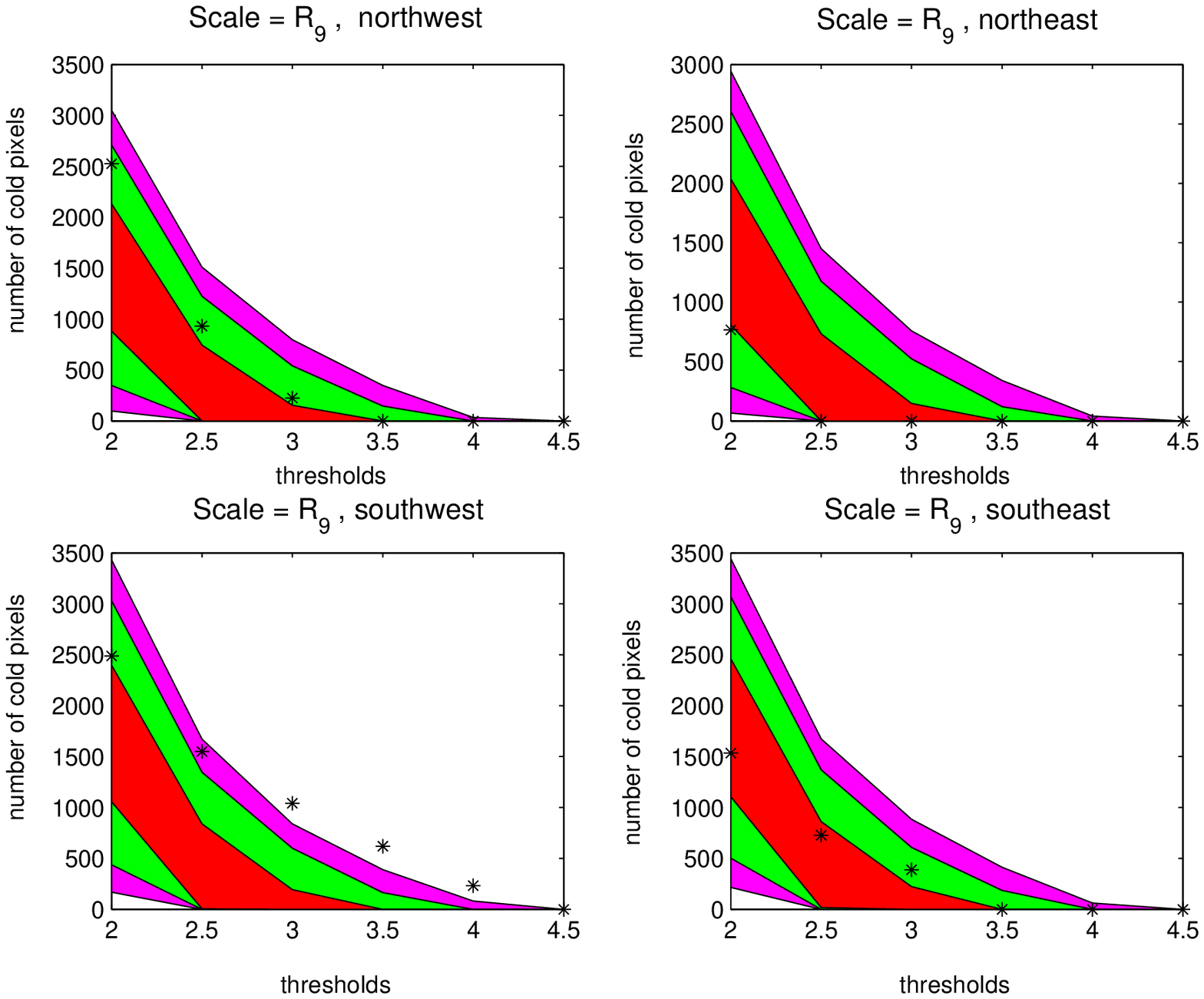}
\end{center}
\caption{Cold area in four regions, scales $R_{8}$ and $R_{9}$. The only non-Gaussian region is 
the southwest. The acceptance intervals are plotted as in Figure ~\ref{fig:all_scales}.}
\label{fig:Area_Cold4}
\end{figure*}
The results presented in Figure ~\ref{fig:Area_Cold4} clearly reveal 
the location of the non-Gaussian signatures. 
The Southwest is the only region where the data lie outside
the acceptance intervals. As expected, the region containing The Spot is not compatible 
with Gaussianity, whereas the other three regions are compatible with a Gaussian behaviour.
Note that although Eriksen et al. 2004b found the ILC weights to have particular values in 
this region, this does not affect our results since we use the combined, cleaned Q-V-W map.

In the next section we quantify the significance of The Spot.
 
\subsection{The Spot}
Our aim in this section is to quantify the probability of finding a spot like the one
found at ($b = -57^\circ, l = 209^\circ$) in a Gaussian, homogeneous and isotropic 
random field, and to 
check whether the map without The Spot is compatible with Gaussianity or not.
\begin{figure*}
\begin{center}
\includegraphics[width=8cm]{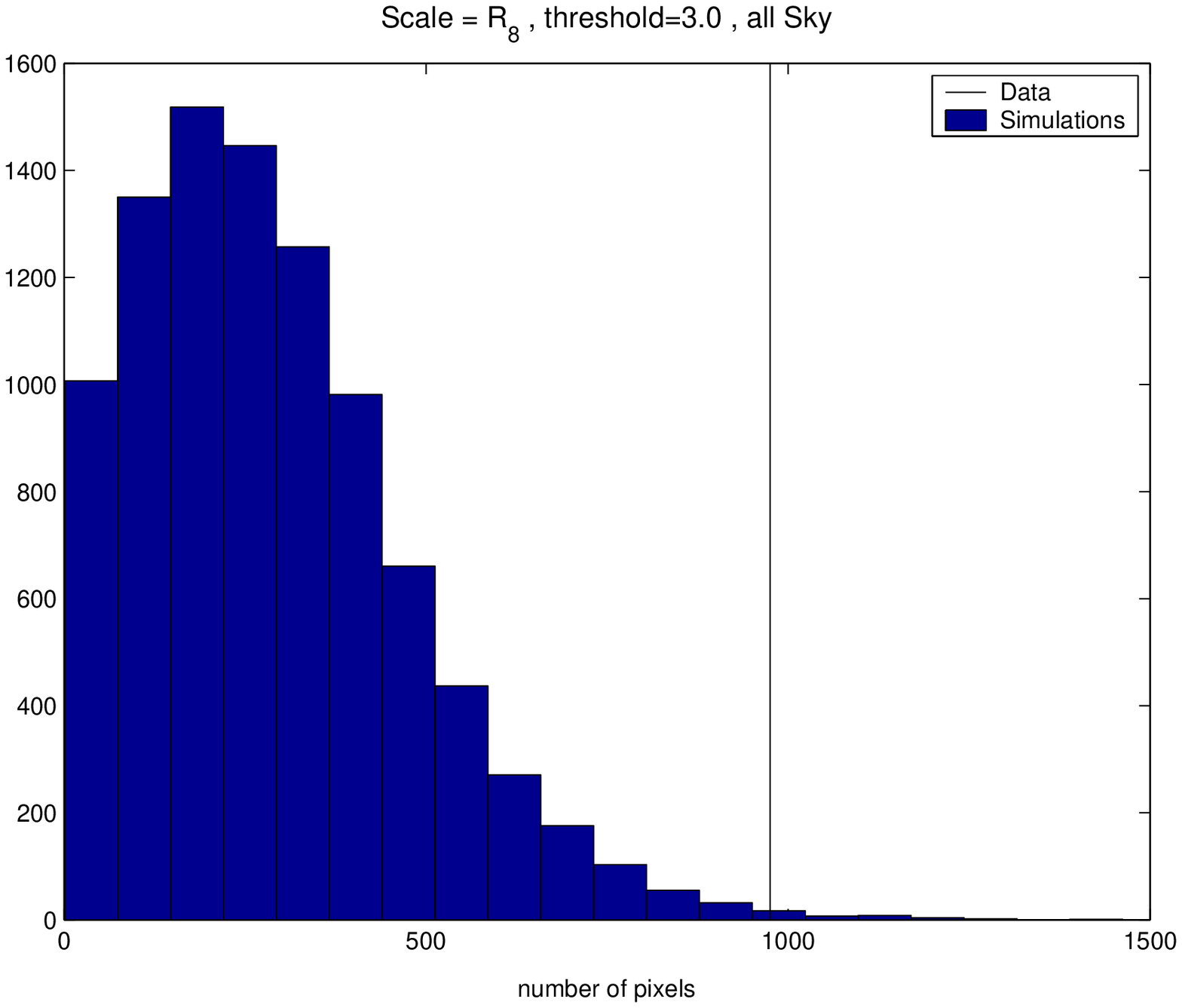}
\includegraphics[width=8cm]{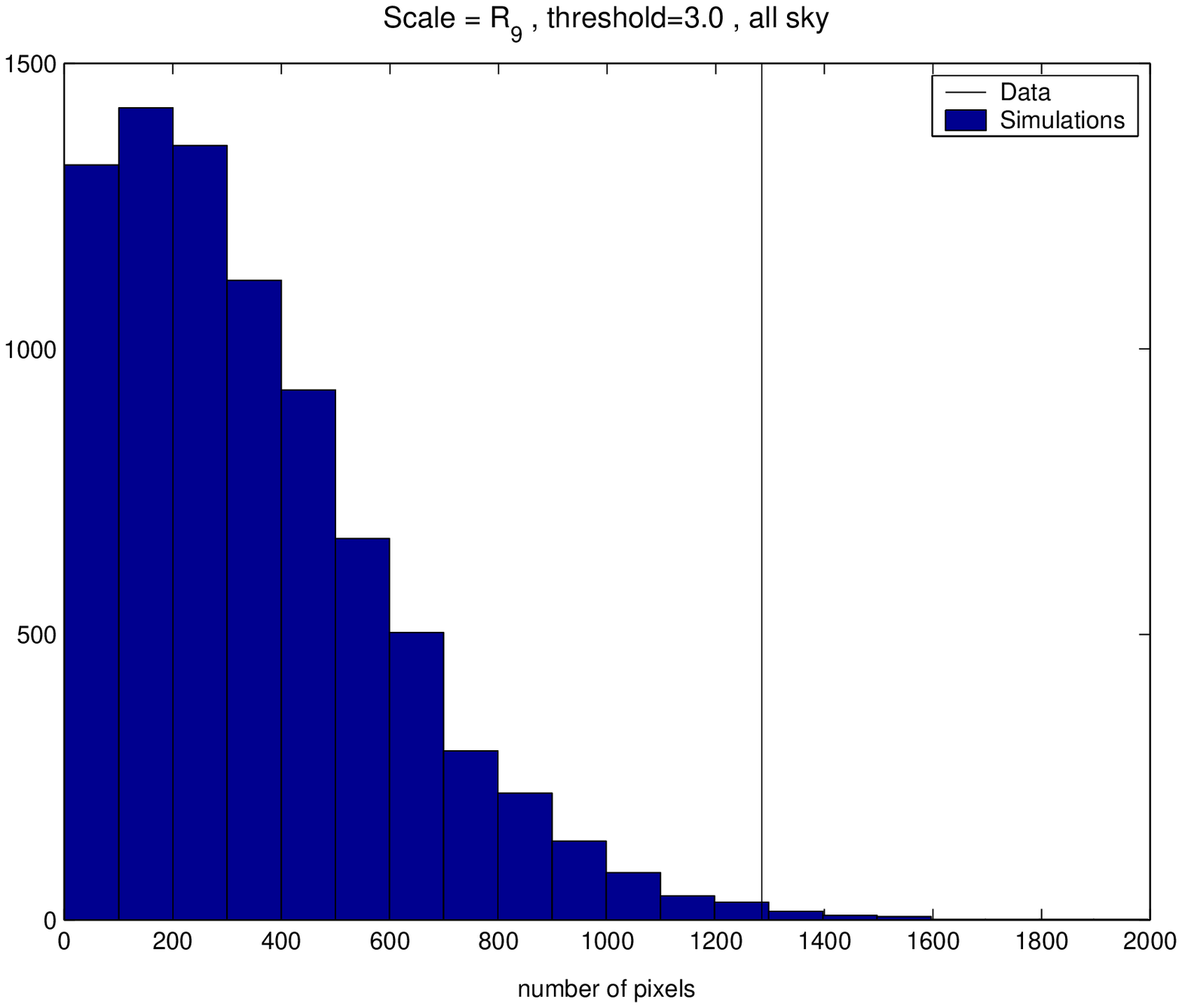}
\end{center}
\caption{Histogram of all biggest cold spots, threshold 3.0, scales $R_{8}$ and $R_{9}$. 
The vertical line represents The Spot.}
\label{fig:Histbig}
\end{figure*}
Comparing The Spot with the biggest cold spot of each simulation we can estimate the 
probability of observing such a spot for the Gaussian model.  
We show the histograms with the biggest cold spot of each simulation
at threshold 3.0 and scales $R_{8}$ and $R_{9}$ in Figure ~\ref{fig:Histbig}.
\begin{table}
   \begin{center}
         \begin{tabular}{|c|c|c|}
	 \hline
	 Scale & threshold & probability \\
	 $R_{8}$ & 3.0 & 0.34\% \\
	         & 3.5 & 0.32\% \\
	         & 4.0 & 0.41\% \\
	         & 4.5 & 0.65\% \\
	 \hline
	 $R_{9}$ & 3.0 & 0.38\% \\
	         & 3.5 & 0.21\% \\
	         & 4.0 & 0.18\% \\
	         & 4.5 & 0.22\% \\
	 \hline
      \end{tabular}
      \caption{Upper tail probabilities of having The Spot under a Gaussian model, at different scales
               and thresholds. All probabilities are smaller than 1\%.}
      \label{table:Big spots}
    \end{center}
\end{table}
The results for all thresholds are summarized in Table ~\ref{table:Big spots}.
Note that some simulations do not have any spots at high thresholds
and hence they do not appear in the histograms but we take them into account to estimate 
the probabilities. 
All probabilities are below 0.7\%. The lowest value is 0.18\% and implies a non-Gaussian 
detection at the 99.82\% level.

At this point we can make the hypothesis that the data could be explained as the sum 
of a Gaussian, homogeneous and isotropic random field, plus a non-Gaussian
spot which is not generated by this field.
With the purpose of checking our hypothesis, we have compared the cold area of 
data and simulations, at scales $R_{8}$, $R_{9}$ and thresholds 3.0, 3.5 
where the data present more than one spot.
\begin{table}
   \begin{center}
         \begin{tabular}{|c|c|c|c|}
	 \hline
	 Scale & threshold & P with Spot & P without Spot \\
	 $R_{8}$ & 3.0 & 0.18\% & 14.79\%\\
	         & 3.5 & 0.28\% & 18.28\%\\
	         & 4.0 & 0.45\% &  - \\
	         & 4.5 & 0.65\% &  - \\
	 \hline
	 $R_{9}$ & 3.0 & 0.39\% & 30.53\%\\
	         & 3.5 & 0.18\% & 17.68\%\\
	         & 4.0 & 0.19\% &  - \\
	         & 4.5 & 0.22\% &  - \\
	 \hline
      \end{tabular}
      \caption{Upper tail probabilities of having the cold area measured 
               in the data, under a Gaussian model
               The third column displays the probabilities considering The Spot
               and the right column shows the probabilities subtracting The 
               Spot from the data. At thresholds 4.0 an 4.5 only The Spot is present, 
	       hence it makes no sense considering probabilities without The Spot.}
      \label{table:Total Area}
    \end{center}
\end{table}
First we have estimated the upper tail probabilities of finding the total cold area 
(including all cold spots), counting
how many simulations present a greater or equal cold area, 
obtaining very low values (see Table ~\ref{table:Total Area}).
If we subtract The Spot from the cold area in the data, and calculate again the probabilities,
we appreciate how the remaining
area is compatible with Gaussianity, since the upper tail probabilities grow in a factor of 100. 
\begin{figure*}
\begin{center}
\includegraphics[width=8cm]{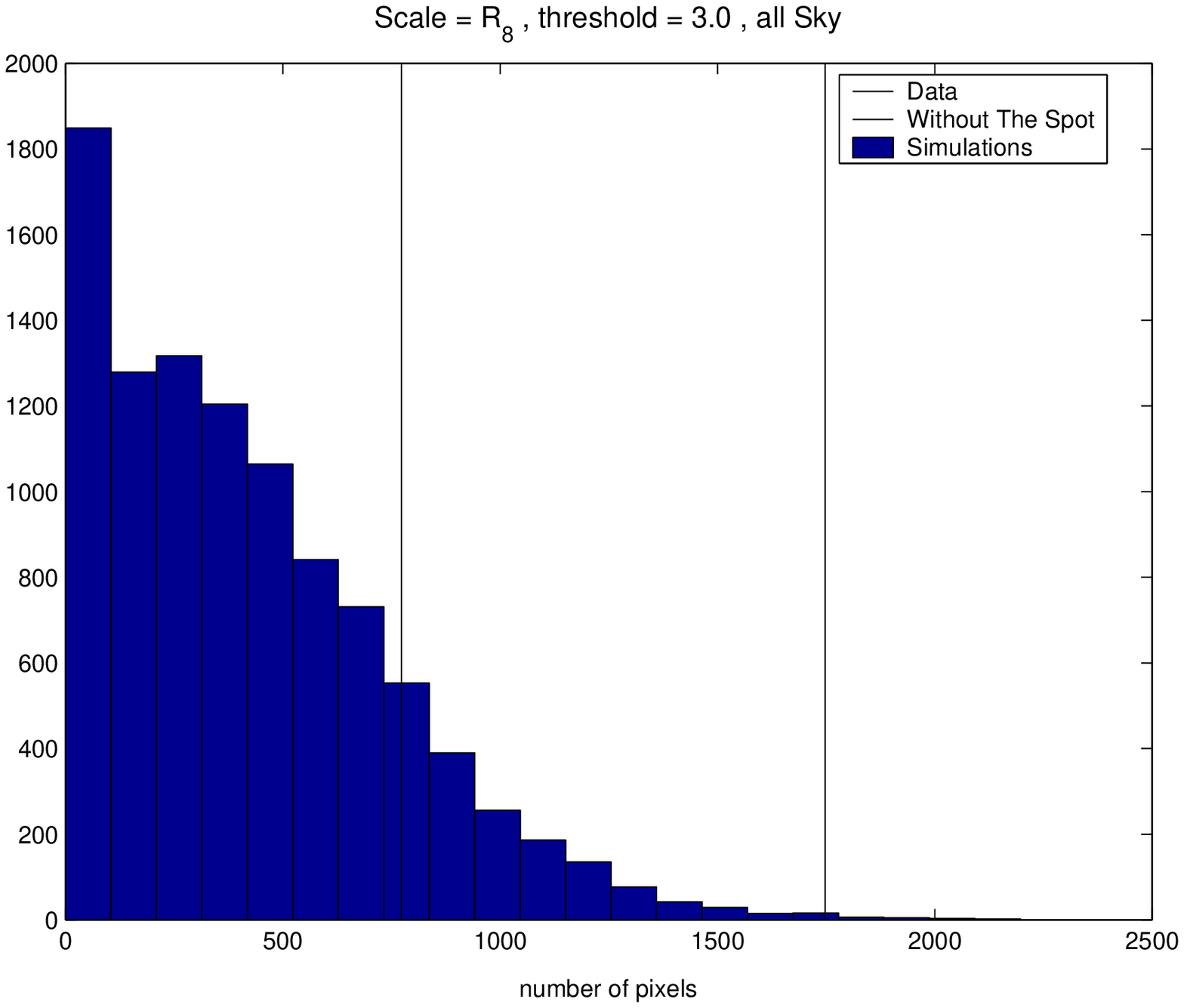}
\includegraphics[width=8cm]{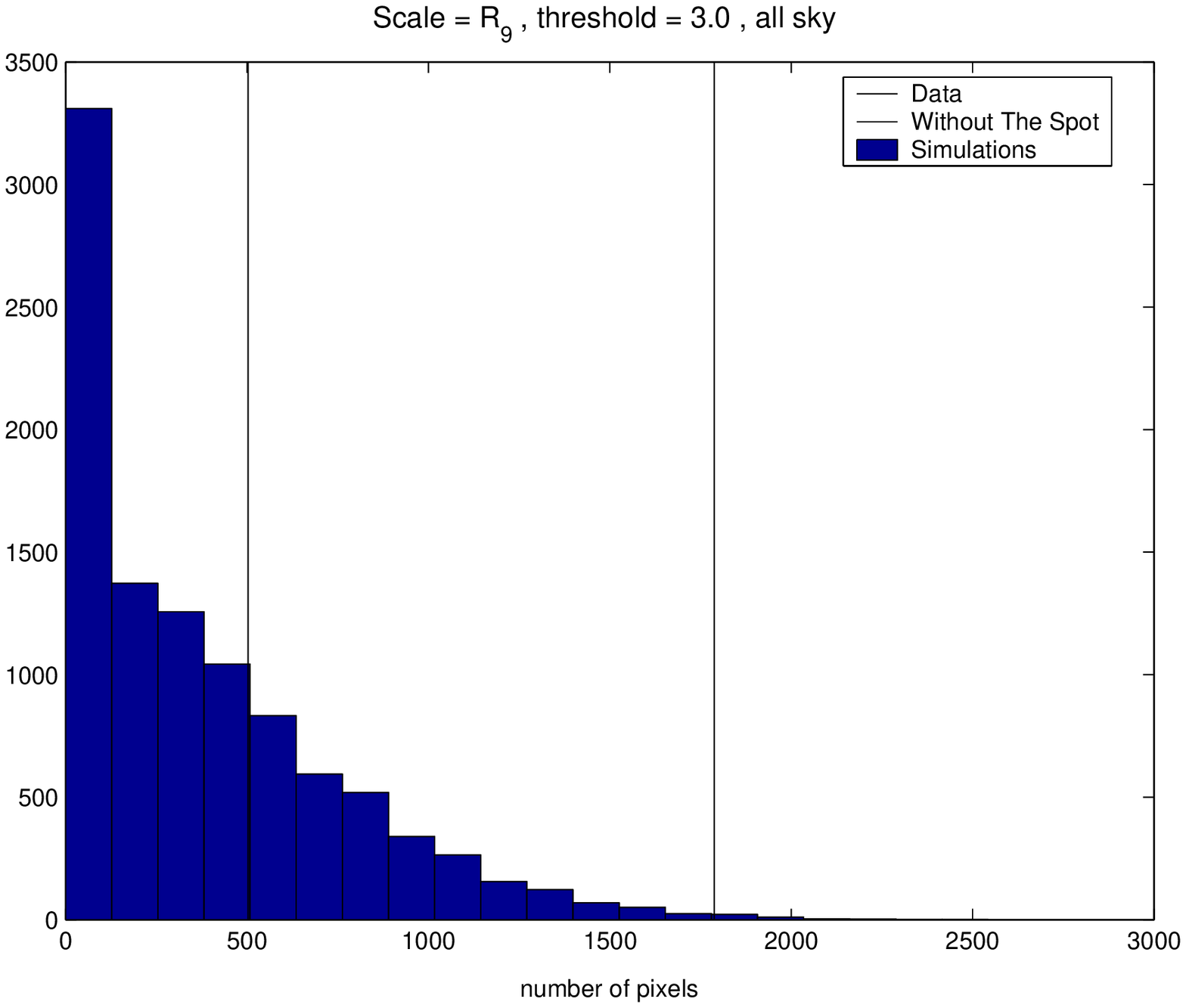}
\end{center}
\caption{Histogram of all cold areas, threshold 3.0, scales $R_{8}$ and $R_{9}$. The data without
The Spot (line on the left) are compatible with Gaussianity, whereas the data with The Spot (line on 
the right), are non-Gaussian with probabilities 99.82\% and 99.61\% at this threshold.}
\label{fig:HistArea}
\end{figure*}
This result is represented in Figure ~\ref{fig:HistArea}. The line on the right hand side 
is the total cold area of the data whereas the one on the left hand side is the area 
remaining after subtracting The Spot. 
The increase in probability can easily be appreciated.
Here we should have considered that we were using fewer
pixels in the data than in the simulations because The Spot has been subtracted only in the data.
But since these pixels represent only about 0.5\% of the total pixels at 
scales $R_{8}$ and $R_{9}$, we can neglect them, without modifying substantially the results.
Hence regarding The Spot as a non-Gaussian outlier, the remaining data are compatible with
Gaussianity.

To finish this section, we want to remark some characteristics of The Spot. 
\begin{table}
   \begin{center}
         \begin{tabular}{|c|c|c|}
	 \hline
	 & Combined WMAP  & \\ 
	 Amplitude ($\mu K$)  & Position $(b,l)$ & area ($\nu = -2$) \\
         -346 & $(-60^\circ,213^\circ)$ & 46 \\
	 -398 & $(-56^\circ,210^\circ)$ & 67 \\
	 -331 & $(-54^\circ,211^\circ)$ & 42 \\
	 -317 & $(-56^\circ,203^\circ)$ & 88 \\
	 \hline
         & Gaussian ($4^\circ$) & \\
	 Amplitude ($\mu K$)  & Position $(b,l)$ & area ($\nu = -2$) \\
	 -73 & $(-57^\circ,209^\circ)$ & 699 \\                     
	 \hline
	 \end{tabular}
	 \caption{Before filtering the combined map, The Spot appears resolved in several spots. At the top we
	   show three characteristic values of the most prominent spots, namely the area below threshold -2, 
	   amplitude and position of the minimum.
	   Filtering with a Gaussian of $4^\circ$ all these spots are convolved, becoming one big spot. The 
	   characteristics of this spot are shown at the bottom.}
	 \label{table:Spot}
   \end{center}
\end{table}
Before convolving the combined map, The Spot appears as several smaller, resolved spots with 
a minimum temperature of $-398\mu K$ at ($b = -56^\circ, l = 210^\circ$). The average value of the four most
prominent spots, listed in table ~\ref{table:Spot} is $-348\mu K$. However, these four spots are not exclusively
responsible for The Spot, since removing them, the mean value of the remaining pixels forming The Spot, is close 
to $-1\sigma$.
We have filtered the combined map with a Gaussian of scale $4^\circ$, 
the characteristic scale given by the SMHW analysis. The Spot appears with an amplitude of $-78\mu K$
at ($b = -56^\circ, l = 210^\circ$) as summarized in Table ~\ref{table:Spot}.

\subsection{Skewness and kurtosis}

At scales $R_{8}$ and $R_{9}$, the kurtosis presented non-Gaussian values in Vielva et al. 2004.
We have repeated the analysis of Vielva et al. 2004, masking The Spot in the wavelet coefficient maps, 
to show to which extent The Spot is responsible for the excess of kurtosis.
The masked pixels, were the pixels of The Spot lying above threshold 3.0 at scale $R_{8}$. 
\begin{figure*}
\begin{center}
\includegraphics[width=8cm]{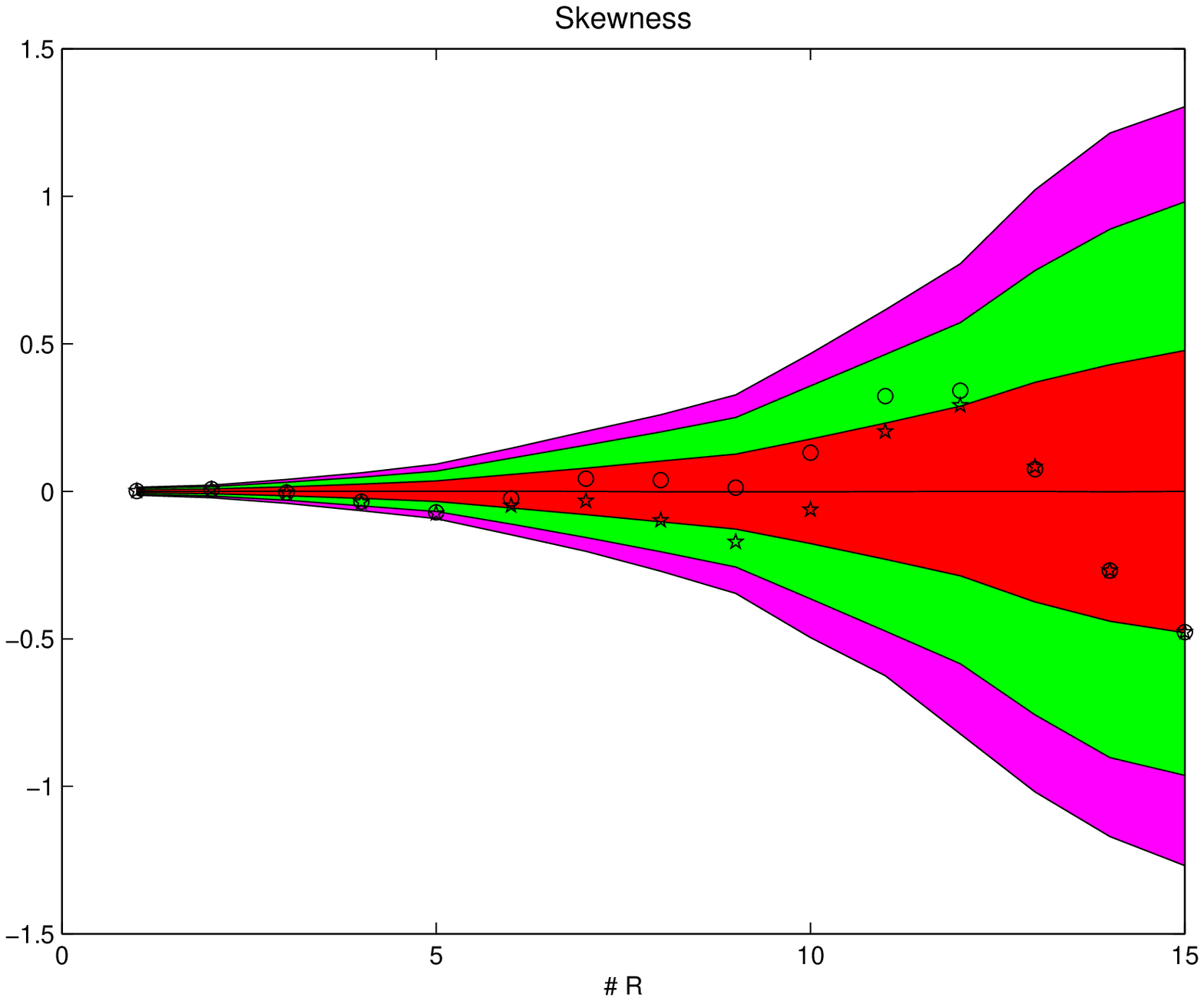}
\includegraphics[width=8cm]{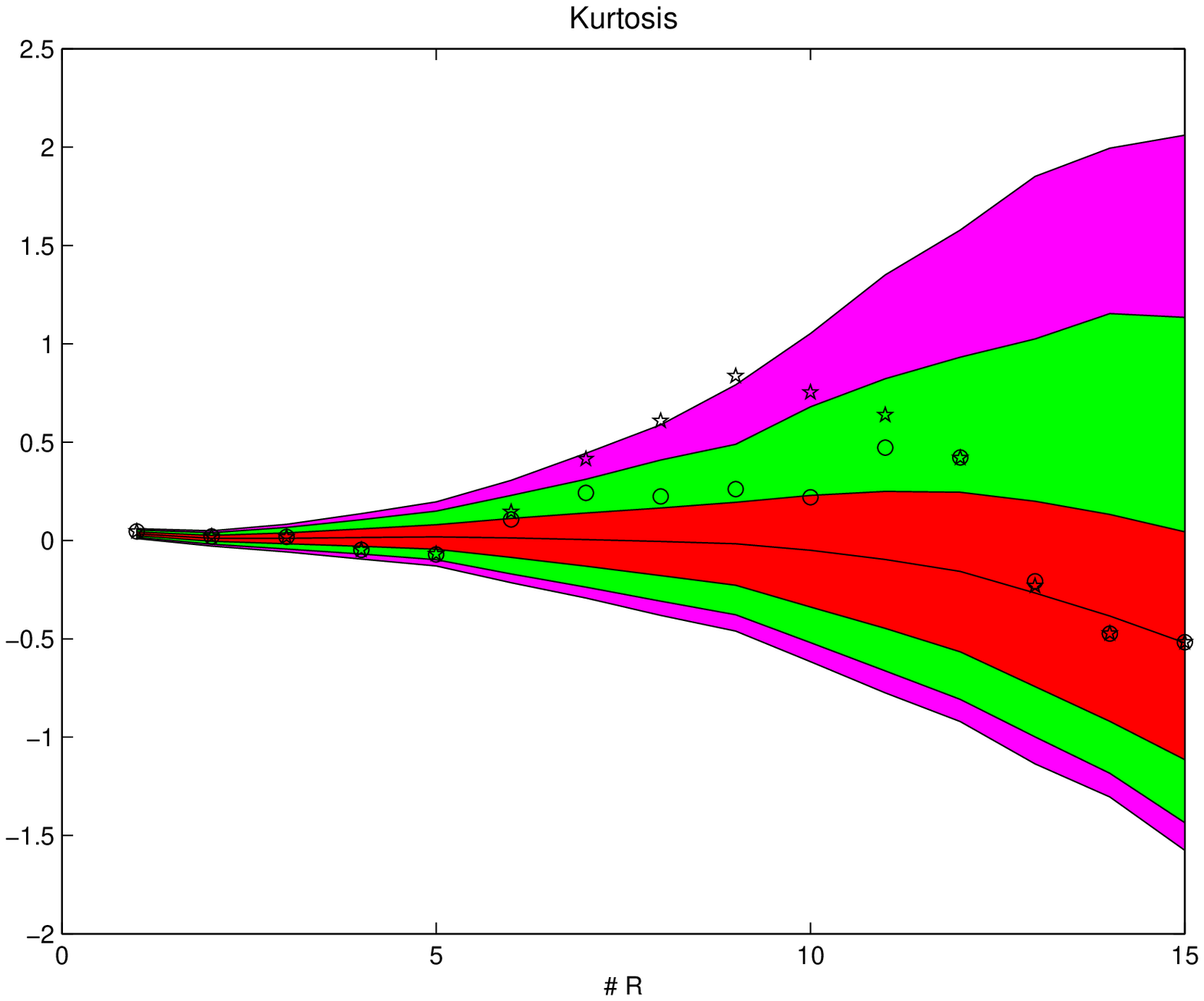}
\end{center}
\caption{This figure shows the values of skewness (left) and kurtosis (right) for all the considered scales.
The combined map values are plotted as stars and as written in Vielva et al. 2004, at scales 
$R_{8}$ and $R_{9}$ they lie outside the three acceptance intervals, which are represented 
as in previous figures. Masking those pixels of The Spot which are above threshold 3.0 at scale $R_{8}$,
we obtain the results represented by circles. The huge decrement of the kurtosis makes the data compatible
with Gaussianity.}
\label{fig:kurtosis}
\end{figure*}
The results are presented in Figure ~\ref{fig:kurtosis}.
The stars represent the data and the circles, the data without the Spot. By subtracting The Spot, the 
decrease of the kurtosis is clearly observed, being now compatible with Gaussianity. 
The acceptance intervals are the same as in Vielva et al. 2004, 
and they were not recalculated masking the pixels of The
Spot in the simulations, since the number of masked pixels is negligible with respect to 
the total number of pixels.
However the decrease in the kurtosis is so huge, that a slight modification of the 
intervals would not affect our
conclusions. The skewness is still compatible with Gaussianity. 
Note that the Spot was masked after convolving with the wavelets. Masking The Spot before convolving,
the decrement of the kurtosis is even higher.
This results show again the Gaussian behaviour of the data without The Spot. 
Hence we can conclude that the excess of kurtosis is exclusively due to The Spot.

\section{Sources of non-Gaussianity}
 
Although Vielva et al. 2004 showed that systematics, foregrounds and variations of the power spectrum
were not responsible for
the non-Gaussian effect shown in the kurtosis, we wanted to check again their influence in
the non-Gaussian results obtained in the present analysis.

\subsection{Systematics}

First we studied the effects of systematics related to instrumental features 
(noise and beam), generating
four sets of 10 simulations. The first two sets are normal simulations with noise and beams.
In the third set we made the same simulations as in the first set but without noise, and in the
fourth set we took the simulations of the second but without beam.    
Comparing the first and the third set, we can see how the noise affects the number of pixels 
in a spot or the total cold area, and comparing sets two and four, we check the influence of 
the beams.
We have compared the spots and the total area of these sets at scales $R_{8}$ and $R_{9}$, and 
threshold 3.0.
In all cases the mean relative variation of the area of one spot was around 2\% and the mean
relative variation of the total area in a simulation around 1\%. This values are negligible
and confirm that noise and beams do not play a significant role in our results,
although the number of considered simulations is not very high. The previous
results of Vielva et al. 2004 support our conclusions.

Another possible source of non-Gaussianity could be the influence of any rare receiver. We
have analyzed the spots detected by the 8 Q-V-W receivers, 
Q1, Q2, V1, V2, W1, W2, W3 and W4, independently.
\begin{figure*}
\begin{center}
\includegraphics[width=8cm]{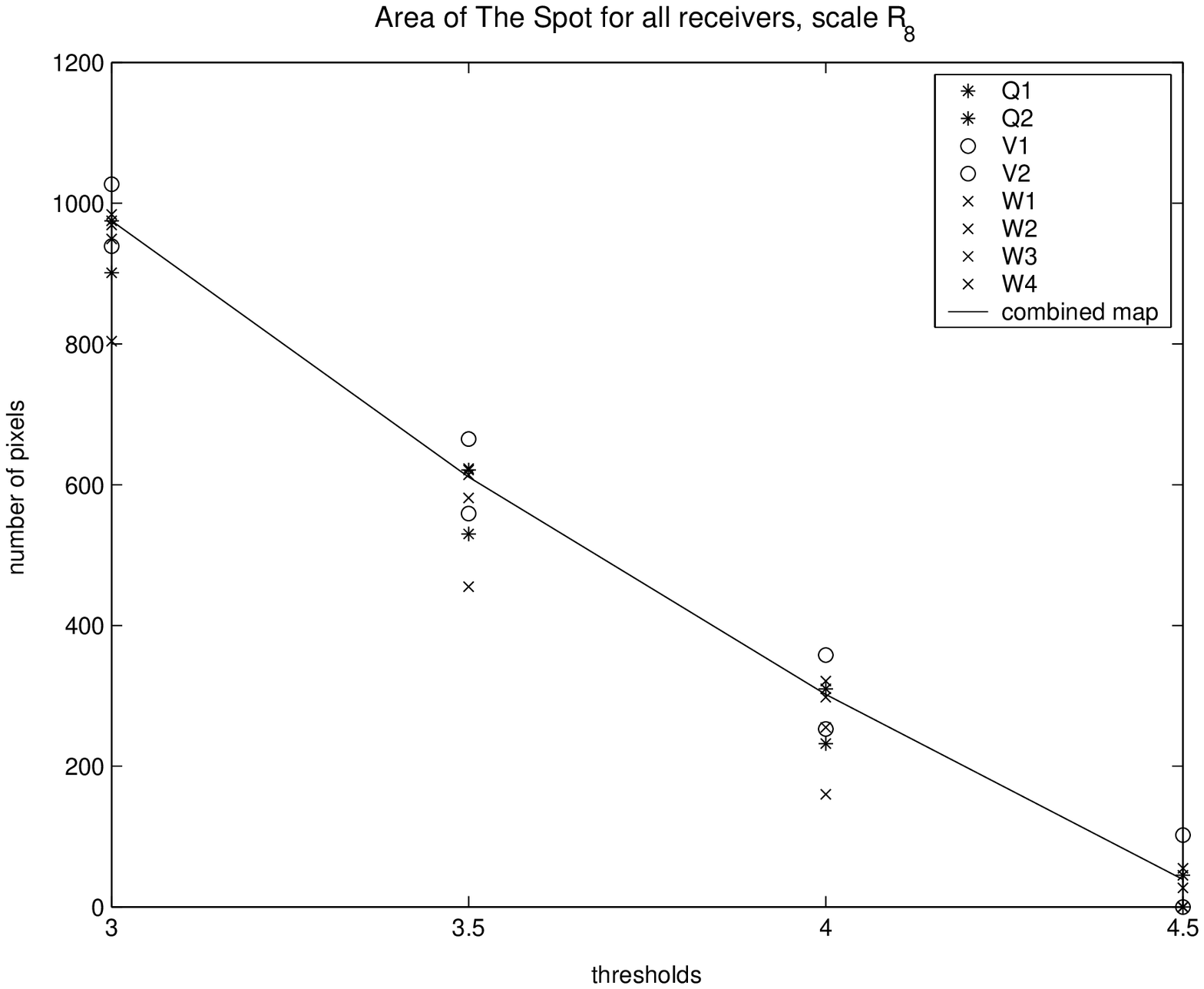}
\includegraphics[width=8cm]{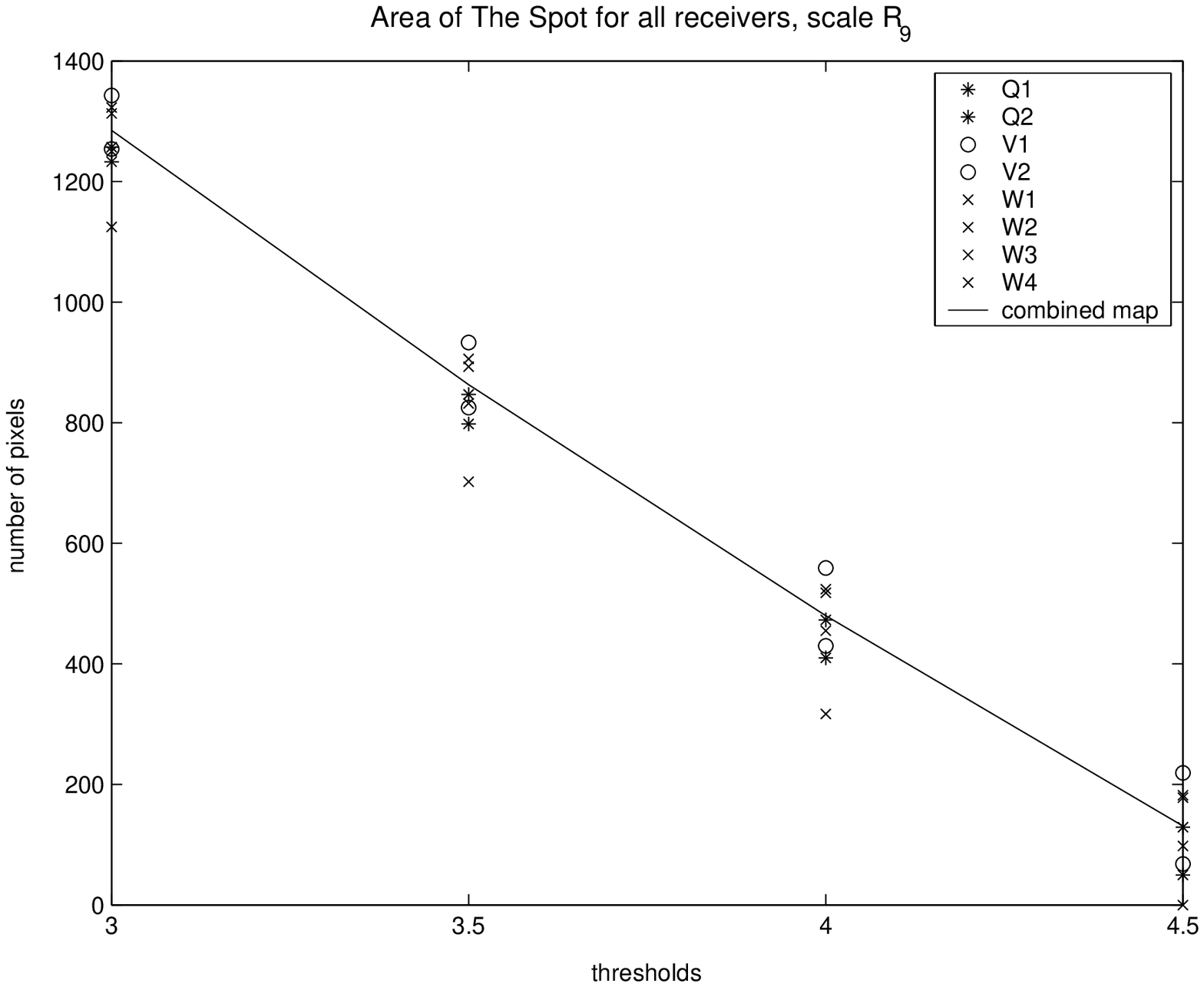}
\end{center}
\caption{Area of The Spot for all receivers, at threshold 3.0 and scales $R_{8}$ and $R_{9}$.
         The line represents the combined map values whereas the other symbols denote the 
         different receivers. The Spot is detected in all receivers, hence no rare receiver is 
         generating the non-Gaussian signal.}
\label{fig:channels}
\end{figure*}
The results for scales $R_{8}$ and $R_{9}$ are plotted in Figure ~\ref{fig:channels}.
Although W2 detects less pixels than the other receivers, all of them detect The Spot, and
are close to the line representing the values of the combined map. 
Hence we verify that our detection is not due to any deficient receiver.

\subsection{Foregrounds}

Non-Gaussianity can be generated by foregrounds due to synchrotron, 
free-free and thermal dust emissions. All foregrounds show a clear
frequency dependence and hence if our Spot is generated by foregrounds its area should also be 
frequency dependent. 
\begin{figure*}
\begin{center}
\includegraphics[width=8cm]{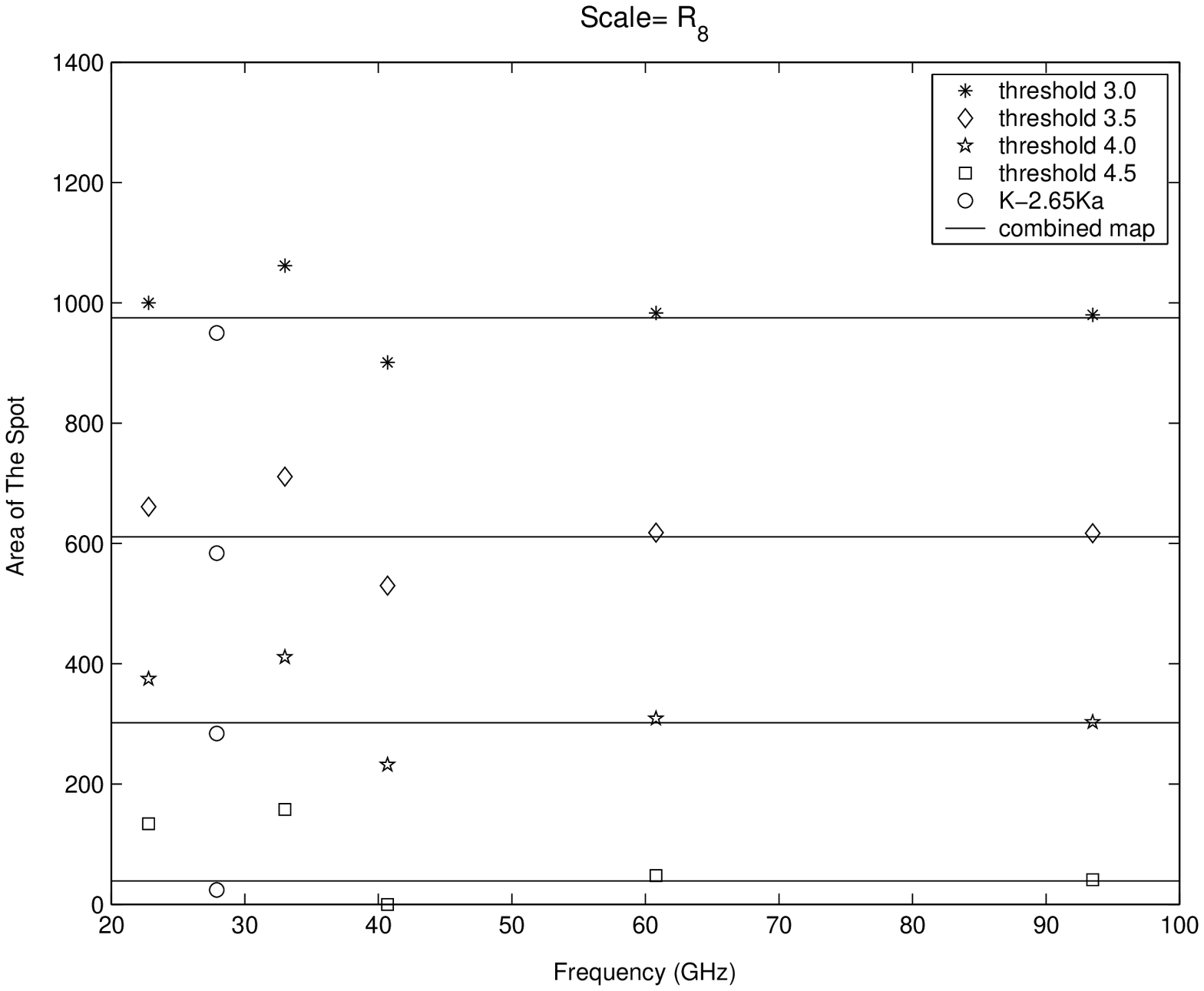}
\includegraphics[width=8cm]{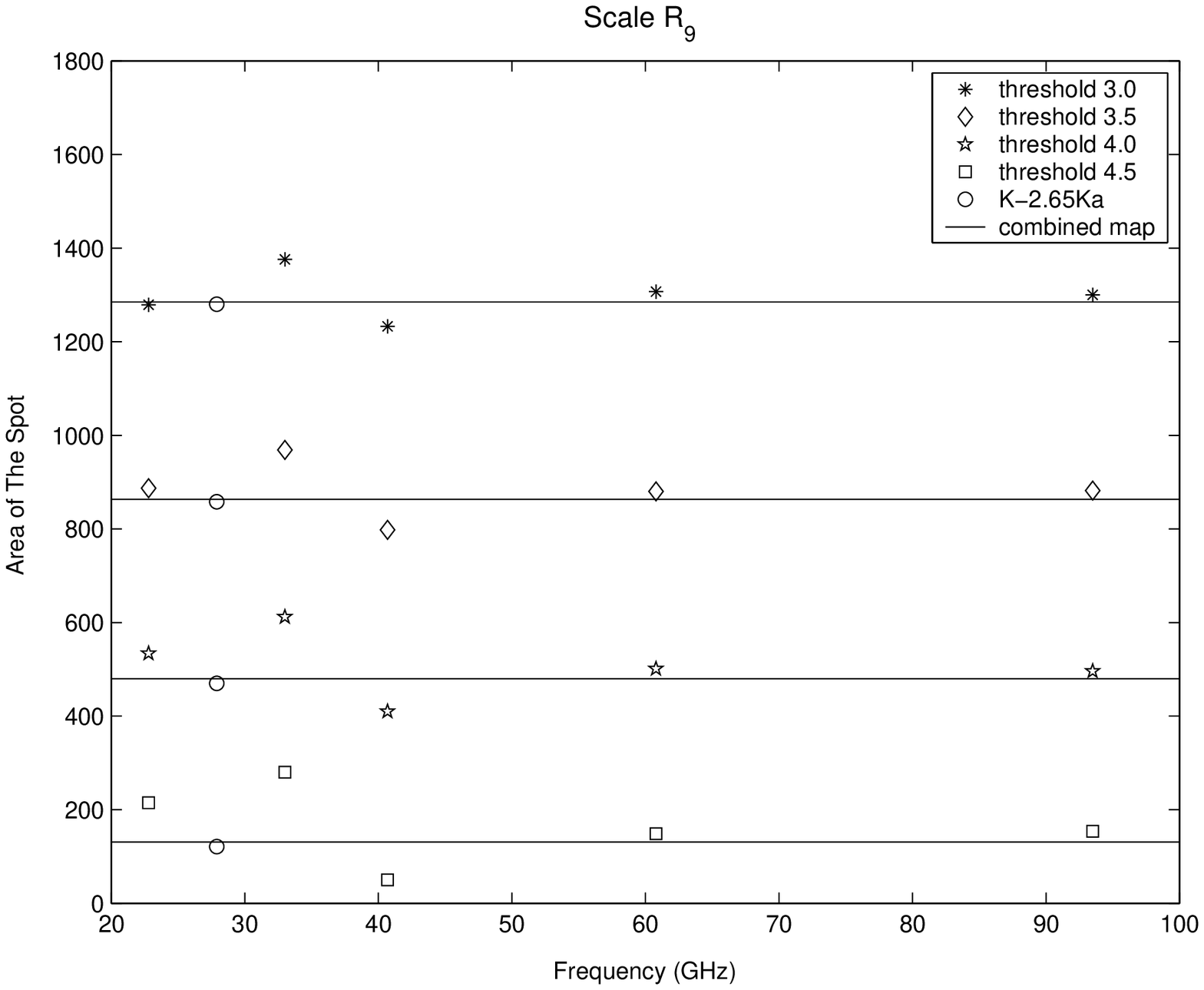}
\end{center}
\caption{Frequency dependence of the area of The Spot, at thresholds over 3.0
         and scales $R_{8}$ and $R_{9}$. The lines represent the combined map values. From top to bottom
	 thresholds 3.0 (asterisks), 3.5 (diamonds), 4.0 (stars) and 4.5 (squares) are plotted. The circles
         are the values for the map $K-2.65 Ka$, where the synchrotron emission should be canceled. These circles are
         close to the combined map values, hence synchrotron emission is not the cause of The Spot.}
\label{fig:freq}
\end{figure*}
Therefore, in Figure ~\ref{fig:freq} we have compared the area of The Spot for each channel, namely 
K(22.8 GHz), Ka(33.0 GHz), Q(40.7 GHz), V(60.8 GHz) and W(93.5 GHz) at scales $R_{8}$ and $R_{9}$ 
and thresholds over 3. The horizontal lines denote the combined map values, whereas 
the other symbols correspond to the area of The Spot at different thresholds for each channel.

The first two channels are not foreground corrected and so they may not
match the results of the other channels and of the combined map. 
In fact as can be seen in Figure ~\ref{fig:freq} both channels deviate slightly from the 
horizontal lines, representing the combined map values. 
Since these channels are not foreground corrected, we can attribute the deviation to the
synchrotron radiation, which dominates at these frequencies. The synchrotron emission is 
expected to grow a factor 2.65 from 33 GHz to 
23 GHz,\footnote{A power law is assumed for the frequency dependence of 
the synchrotron emission: $T_{syn}(\nu) \propto T_{syn}(\nu_0) 
\Big(\nu/\nu_0\Big)^{-2.7}$, as proposed by Bennett et al. (2003b).} hence if we subtract 2.65 times
the Ka-map from the K one, we get rid of the contaminating emission. Considering 
Figure ~\ref{fig:freq},
we confirm that the circles corresponding to K-2.65Ka are very close to the combined map
values.

These results support the conclusions reached in Vielva et al. 2004 regarding the influence
of foregrounds in the non-Gaussian detection
The independence of the amplitude of The Spot with frequency, was already shown in the mentioned paper.

\subsection{Power spectrum dependence}

Since several anomalies and asymmetries have been found in the low multipoles of the power spectrum,
we should discuss to which extent the result depends on the power spectrum. 
We have used the best fit WMAP power spectrum to perform the 10000 Gaussian simulations, 
but the uncertainties in the cosmological parameters and hence in the power spectrum could 
affect our results.
We have performed three sets of 50 simulations. 
The first set corresponds to the best fit WMAP power spectrum. The other
two sets were generated with power spectra differing by $\pm 1\sigma$ from the best fit WMAP 
power spectrum.
One corresponds to the 'lower limit' power spectrum, obtained subtracting the $1\sigma$ error estimated
by the WMAP team, and the other one to the 'upper limit' power spectrum, adding the 
mentioned $1\sigma$ error to the best fit spectrum. Comparing the first set with the two others,
we can study how the spots are affected by the choice of different power spectra.

At scales $R_{8}$ and $R_{9}$, and threshold 3.0 the mean relative variation of the area of a 
particular cold spot is much lower than 1\%. Therefore, after these negligible variations, the significance 
of The Spot remains unchanged. 
Also Vielva et al. 2004 found a negligible power spectrum dependence of the acceptance intervals for 
the kurtosis.
Hence the choice of different power spectra does not affect significantly our results.

\subsection{Intrinsic anisotropies}

Once we have discarded systematics and foregrounds as the cause of our detection, other 
sources have to be considered. For instance the Sunyaev-Zeldovich effect could produce
a cold spot. This effect occurs in clusters, when high energetic electrons collide with 
CMB photons originating a decrement of the temperature in our range of frequencies. Here arise
two problems, namely the angular scale and the amplitude of The Spot. 
The non-Gaussianity is found at scales around $4^\circ$ implying a size of  
about $10^\circ$ on the sky. Observing in real space 
this region, The Spot appears resolved in several smaller very cold spots with a minimum temperature
around $-398\mu K$. We have looked for any extragalactic object which could cover this
angular scale at coordinates near ($b = -57^\circ, l = 209^\circ$). 
We found a group of galaxies belonging
to the local supercluster at a distance of about 20 Mpc subtending a similar angle on the sky. 
Some of these galaxies match the resolved spots observed in the data. 
However the mass and temperature of the gas necessary to reach the amplitude of our Spot, are similar
to the values found in rich clusters and therefore much higher than the amounts estimated 
for groups of galaxies (see Taylor et al. 2003). 
We would need a big cluster, such as the Coma one, to reach such an amplitude and it 
should be near enough to cover 10 degrees on the sky. 
In the neighborhood of  ($b = -57^\circ, l = 209^\circ$) no such object is found.
This is in agreement with the WMAP results on foregrounds (Bennett et al 2003b) where the
Sunyaev-Zeldovich effect was found to be negligible except for the most prominent nearby cluster,
Coma, observed with a signal to noise ratio of $\approx 2$. 
Even more, we have also considered the ACO catalogue and the All-Sky ROSAT maps
(Snowden et al. 1997) at 0.1, 1.2 and 2 keV and neither any ACO cluster nor 
any special X-ray emission was found at the The Spot position.
However, the ROSAT maps present some particular problems:
the brightest point sources, as well as a large fraction of 
clusters, have been removed from it. In addition, some small fractions 
of the sky were not observed and, unfortunately, one of them is very 
close to our object.

Another possible source is the Rees-Sciama effect, (Mart\'{\i}nez-Gonz\'{a}lez et al. 1990, 
Mart\'{\i}nez-Gonz\'{a}lez \& Sanz 1990).
An extremely massive and distant superstructure would be a clear candidate to cause a
cold and big secondary anisotropy.
Even topological defects like global monopoles or textures (Turok \& Spergel 1990) could have cooled
the CMB photons, to produce such spot. Cosmic strings have characteristic 
scales around arcminutes, hence we do not 
expect them to be behind this non-Gaussian detection.

\section{Conclusions}

Motivated by the non-Gaussianity found in the WMAP 1-year data using
the SMHW, we have performed an analysis of the spots in the SMHW
coefficients map, aimed to locate possible contributors in the sky.  
An extremely cold and big spot is detected. This spot (\emph{The Spot}) is
seen in the SMHW coefficients at scales
around $4^\circ$ (implying a size of around $10^\circ$ on the sky) and at
Galactic coordinates $b=-57^\circ, l=209^\circ$. The probability of
having such spot for a Gaussian model is of only $\approx 0.2\%$, which
implies that, if intrinsic, The Spot has not been originated by
primary anisotropies in the standard scenario of structure formation
since standard inflation predicts Gaussian fluctuations in the matter
energy density and therefore in the CMB temperature fluctuations.
When this spot is not considered in the analysis the rest of
the data seem to be consistent with Gaussianity.

In order to identify the source of The Spot we have performed several
tests related to systematic effects and foregrounds. We have checked
that uncertainties in the noise or in the beam response have a
negligible effect in our results at the relevant wavelet scales.  
Looking at the maps corresponding to the different receivers,
we see a clear consistency in the area, amplitude and position of The
Spot. Hence our detection is not due to any deficient receiver.
In relation to the possible foregrounds contribution, we have
looked for possible frequency dependences in the amplitude and area of
The Spot. Again both quantities show a nice consistency with a
constant line in the range from 23 to 94 GHz. Whereas the Galactic
foregrounds show a very different frequency dependence with respect to
the constant behaviour, the SZ effect does not separate much
from it in that frequency range. 
A comparable spot could be produced either by the Coma cluster at a much closer
distance, or by several rich clusters at the actual distance of Coma.
We have checked that no nearby rich cluster of galaxies is
located in the position of The Spot.
  
Finally, intrinsic fluctuations cannot be rejected as the source of
The Spot. In particular, a massive and distant super-structure could
in principle produce a decrement as the one observed through the
Rees-Sciama effect (Mart\'\i nez-Gonz\'alez and Sanz 1990). This
massive structure (of order of at least 
$10^{16} M_\odot$) should be placed far enough
because otherwise it would have been detected previously.
Alternatively, more speculative possibilities are topological defects
(monopoles or textures) or non-standard inflationary scenarios.
Even more, a combination of secondary anisotropies with primary ones, 
cannot be rejected as the source of our non-Gaussian spot. For instance a
possibility could be a combination of the Sunyaev-Zeldovich effect with a Sachs-Wolfe plateau.

\section*{acknowledgments}
Authors kindly thank J. M. Diego for very useful discussions about the 
Sunyaev-Zeldovich effect in the local galaxy distribution, C. Hern\'andez-Monteagudo for
helpful comments about the ACO catalog and R. Durrer for comments on topological defects.
MC thanks Spanish Ministerio de Educacion Cultura y Deporte (MECD) for a predoctoral FPU fellowship.
PV acknowledges support from IN2P3 (CNRS) for a post-doc fellowship
We acknowledge partial financial support from the Spanish MCYT project ESP2002-
04141-C03-01.
We kindly thank IFCA for providing us with its Grid Wall cluster to generate and 
analyze the WMAP simulations.
We thank the RTN of the EU project HPRN-CT-2000-00124 for partial financial
support.
We acknowledge the use of the Legacy Archive for Microwave Background Data 
Analysis (LAMBDA). Support for LAMBDA is provided by the NASA Office of Space 
Science.
This work has used the software package HEALPix (Hierarchical, Equal
Area and iso-latitude pixelization of the sphere,
http://www.eso.org/science/healpix), developed by K.M. G{\'o}rski,
E. F. Hivon, B. D. Wandelt, J. Banday, F. K. Hansen and
M. Barthelmann.
We acknowledge the use of the software package CMBFAST
(http://www.cmbfast.org) developed by U. Seljak and M. Zaldarriaga.

\end{document}
\end